\documentclass[12pt]{article}
\usepackage{epsfig}
\usepackage{amsfonts}
\usepackage{amsmath}
\usepackage{amssymb}

\newcommand{\simle}{\hspace*{0.2em}\raisebox{0.5ex}{$<$}
     \hspace{-0.8em}\raisebox{-0.3em}{$\sim$}\hspace*{0.2em}}

\newcommand{\bt}{\beta}

\newcommand{\Fmu}{F_{\mu\nu}}

\newcommand{\slashT}{\slash\hspace{-0.4em}T}

\newcommand{\slashP}{\slash\hspace{-0.6em}P}

\newcommand{\slashPT}{\slash\hspace{-0.6em}P\slash\hspace{-0.5em}T}
\newcommand{\slashPTsub}{\slash\hspace{-0.45em}P\slash\hspace{-0.4em}T}

\newcommand{\Nb}{\bar N}
\newcommand{\Fp}{F_\pi}

\newcommand{\tb}{\bar \theta}

\newcommand{\mpi}{m_{\pi}}
\newcommand{\MQCD}{M_{\mathrm{QCD}}}

\newcommand{\Or}{\mathcal O}

\newcommand{\dslash}[1]{#1 \llap{/\kern-0.5pt}}
\newcommand{\Dslash}[1]{#1 \llap{/\kern+1.2pt}}
\newcommand{\DDslash}[1]{#1 \llap{/\kern+2.3pt}}
\newcommand{\dslashh}[1]{#1 \llap{/\kern+1pt}}
\newcommand{\abs}[1]{|#1|}

\newcommand{\boldtau}{\mbox{\boldmath $\tau$}}
\newcommand{\boldpi}{\mbox{\boldmath $\pi$}}

\begin{document}

\begin{titlepage}

\vspace{2.0cm}

\begin{center}
{\Large\bf Electric Dipole Moments of Light Nuclei\\
\vspace{0.2cm}
{}From Chiral Effective Field Theory}

\vspace{1.7cm}

{\large \bf J. de Vries$^1$, R. Higa$^{1,2}$,  
C.-P. Liu$^3$, E. Mereghetti$^{4,5}$, I. Stetcu$^6$,\\
\vspace{0.2cm}
R. G. E. Timmermans$^1$, and U. van Kolck$^4$}

\vspace{0.7cm}
{\large 
$^1$ 
{\it KVI, Theory Group, University of Groningen,
\\
9747 AA Groningen, The Netherlands}}

\vspace{0.25cm}
{\large 
$^2$ 
{\it Instituto de F\'\i sica, Universidade de S\~ao Paulo, 
\\ 
C.P. 66318, 05389-970 S\~ao Paulo, SP, Brazil}}

\vspace{0.25cm}
{\large 
$^3$ 
{\it Department of Physics, National Dong Hwa University, 
\\
Shoufeng, Hualien 97401, Taiwan}}

\vspace{0.25cm}
{\large 
$^4$ 
\it Department of Physics, University of Arizona,
\\
Tucson, AZ 85721, USA}

\vspace{0.25cm}
{\large 
$^5$ 
\it Ernest Orlando Lawrence Berkeley National Laboratory,\\
University of California, Berkeley, CA 94720, USA}

\vspace{0.25cm}
{\large 
$^6$ 
{\it Department of Physics, University of Washington, Box 351560, 
\\
Seattle, WA 98195-1560, USA}}

\end{center}

\vspace{1.5cm}

\begin{abstract}
We set up the framework for the calculation of electric dipole
moments (EDMs) of light nuclei using the systematic expansion
provided by chiral effective field theory (EFT).
We take into account parity ($P$) and time-reversal ($T$) violation 
which, at the quark-gluon level, originates from
the QCD vacuum angle and dimension-six operators capturing
physics beyond the Standard Model.
We argue that EDMs of light nuclei can be expressed in
terms of six low-energy constants that appear in the 
$P$- and $T$-violating nuclear potential and electric current.
As examples, we calculate the EDMs of the deuteron, the triton,
and $^3$He in leading order in the EFT expansion.
\end{abstract}

\vfill
\end{titlepage}

\section{Introduction}

Permanent electric dipole moments (EDMs) \cite{KhripLam1997,Pospelov:2005pr}
of particles, nuclei, atoms, and molecules provide stringent bounds on sources
of parity ($P$) and time-reversal ($T$) violation beyond the phase of the 
quark-mixing matrix \cite{Kobayashi:1973fv} in the Standard Model (SM).
Experiments in preparation \cite{expts} aim to improve the
current bound on the neutron EDM, $|d_{n}|<2.9\cdot10^{-26}\, e$
cm \cite{dnbound}, by up to two orders of magnitude. At the same
time, proposals \cite{storageringexpts} exist for the measurement
of EDMs of charged particles in storage rings, in particular the proton
--- for which an indirect bound, $|d_{p}|<7.9\cdot10^{-25}\, e$ cm,
has been extracted from the $^{199}$Hg EDM \cite{hgbound} --- and
the deuteron. It may be possible, moreover, to measure in this way
the EDMs of other light nuclei, for example $^{3}\mathrm{He}$
and $^{3}\mathrm{H}$, as well.  A nonzero EDM signal in this
new generation of experiments would be an unambiguous sign of new
physics at energy scales similar to those probed by the LHC,
since the EDMs resulting from the quark-mixing
matrix \cite{McKellar:1987tfPospelov:1994uf} are orders
of magnitude below the current experimental limits.

An important outstanding question is whether it will be possible to
identify from the experimental results the microscopic parity- and
time-reversal-violating ($\slashPT$) source. The SM
contains, apart from the quark-mixing phase and its lepton analog,
also the $\slashPT$ QCD $\tb$ term \cite{'tHooft:1976up}. This interaction
has dimension four and one might expect it to give the main contribution
to hadronic $\slashPT$. However, since the experimental upper limit
on the neutron EDM constrains the vacuum angle
$\tb$ to be unnaturally small~\cite{Bal79,Cre79},
$\tb\simle10^{-10}$, non-SM contributions from higher-dimensional
$\slashPT$ sources can be relevant, or even dominant. These higher-dimensional
operators originate from an ultraviolet complete theory at a high energy
scale $M_{\slashT}$ beyond the electroweak scale. The first such effective
$\slashPT$ operators one can write down have effective dimension
six~\cite{dim6origin,Weinberg:1989dx,Grzadkowski:2010es,RamseyMusolf:2006vr},
\textit{viz.} the quark EDM (qEDM), the quark and gluon chromo-electric
dipole moments (qCEDM and gCEDM, respectively), and two four-quark
(FQ) interactions.

It is difficult to calculate hadronic properties directly
from a Lagrangian written in terms of quark and gluon fields.
Still, long-distance
effects of the strong interactions can be described independently
of assumptions about the dynamics of QCD if one uses the low-energy
effective field theory (EFT) known as chiral perturbation theory ($\chi$PT)
\cite{weinberg79,HBChPT} (for reviews see Refs. \cite{Weinbergbook,ulfreview}),
which translates microscopic operators into operators that contain
nucleon, pion, and photon fields. After the translation, one is able
to calculate hadronic properties directly from the effective Lagrangian.
For the nucleon EDM (and its associated form factor) stemming from
the dimension-four $\tb$ term, this method was first used in Refs.
\cite{Bal79,Cre79}, and later extended in the context of $SU(2)\times SU(2)$
\cite{Thomas,BiraHockings,BiraEmanuele,Mer11} and $SU(3)\times SU(3)$
\cite{su3,Ottnad} $\chi$PT. In this way it is possible to establish
a link to $PT$-conserving ($PT$) 
interactions \cite{BiraHockings,BiraEmanuele,Mer11}.

This approach can be generalized to include the dimension-six sources
\cite{Vri11a,Mer11,dim6}. Different $\slashPT$ sources at the quark-gluon
level produce different hadronic interactions. The effective chiral
Lagrangian includes not only interactions that stem from spontaneous
chiral symmetry breaking and are therefore chiral invariant, but also
interactions that break chiral symmetry in the same way as 
chiral-symmetry-breaking
operators at the QCD level. Thus, while they all break $P$ and $T$,
the dimension-six operators break chiral symmetry differently from
each other and from the $\tb$ term. Given enough observables it should
be possible to separate the various $\slashPT$ sources on the basis
of the hadronic interactions that they generate.

Recently it was argued \cite{Vri11b} that a measurement of the deuteron
EDM in combination with the neutron or proton EDM could partially
separate the fundamental $\slashPT$ sources. A measurement of the
deuteron EDM significantly larger than the nucleon EDM would point
toward new physics in the guise of a quark chromo-EDM. The calculation
was based on a perturbative-pion approach \cite{KSW,nEFT} to nuclear EFT,
which assumes that pion exchange can be treated in perturbation theory. A major
advantage of this approach is that analytical results can be obtained.
On the other hand, such a framework is applicable only below the momentum scale
($\sim300$ MeV) at which one-pion-exchange
(OPE) becomes significant. This is the case
for nuclei where the binding momentum per nucleon is small compared
to the pion mass, but even then the size of uncertainties is set by
the inverse of the relatively low energy scale.

Our goal in this article is to provide a framework for the calculation
of the EDMs of light nuclei using chiral EFT with nonperturbative
OPE \cite{weinberg,vanKolck,morethanweinberg,nEFT}. By treating OPE
nonperturbatively, the EFT gets extended to higher momenta and thus
denser nuclei, and convergence improves. The fact that nuclear binding
momenta are small in the typical scale of QCD ($\sim1$ GeV)
is sufficient for a general power counting that is able to estimate
which hadronic interactions are dominant for each fundamental $\slashPT$
source. The $\slashPT$ potential has been derived 
previously \cite{TVpotential},
and here we obtain the associated $\slashPT$ currents. As explicit
examples we consider the EDMs of the deuteron ($^{2}$H), the triton
($^{3}$H), and the helion ($^{3}$He) %
\footnote{Similar calculations are being carried out by a group at FZ J\"ulich
\cite{juelich}.}.

The EDMs of the deuteron \cite{SFK84,Avi85,Khrip,Liu04,Afn10} and
helion \cite{Avi86,Ste08} have been investigated previously within
traditional meson-exchange frameworks. In the most comprehensive studies
\cite{Liu04,Ste08} one started from {}``realistic'' nuclear-force models
and a general $\slashPT$ nucleon-nucleon ($N\! N$) 
interaction \cite{Liu04,Her66}.
The EDMs were expressed in terms of three $\slashPT$ non-derivative
pion-nucleon interactions, which are often assumed to be of similar
size and dominate the EDMs, and in addition short-range $\slashPT$
interactions due to the exchange of heavier mesons were included.
The major advantage of a chiral EFT framework is that it has a direct
link to QCD and exploits the chiral properties of the fundamental
$\slashPT$ sources. Moreover, the power-counting scheme allows a
perturbative expansion such that the theoretical uncertainties can
be estimated and the results can be improved systematically.

When the chiral-symmetry properties of the dimension-four and dimension-six
operators are considered new insights are in fact obtained
\cite{BiraHockings,BiraEmanuele,Vri11a,Mer11,Vri11b,dim6}.
At leading order, only two of the three $\slashPT$ pion-nucleon interactions
contribute. Moreover, there are in general at the same order more
contributions, \textit{viz.} short-range contributions to the neutron
and proton EDMs and two $\slashPT$ $N\! N$ contact
interactions. As we will demonstrate below,
the EDMs of light nuclei can be expressed in terms of these six $\slashPT$
parameters, or low-energy constants (LECs).
(Other $\slashPT$ moments, such as the deuteron magnetic quadrupole
moment, depend in addition on $\slashPT$ pion-nucleon-photon interactions
\cite{Vri11b}.)
For three of the four $\slashPT$ sources, only a subset of these
six LECs is in fact needed. Each LEC can in principle be calculated
from the underlying $\slashPT$ source using an explicit solution
of QCD at low energies, for example through lattice simulations. Compared
to nucleons, the EDMs of light nuclei can give crucial complementary
information about the fundamental $\slashPT$ source. However, the
conventional
assumption that the three $\slashPT$ pion-nucleon interactions can
cover the whole range of nuclear EDMs is oversimplified.

For the $PT$ potential we use here realistic phenomenological potentials
\cite{Sto94,Wir95,eftinspiredNN}.
This {}``hybrid'' approach \cite{hybrid} is justified whenever
there is little sensitivity to the details of short-range physics,
since such realistic potentials all include the long-range pion exchange
that appears in chiral EFT at LO. Such an approach has been tested
successfully for other observables \cite{nEFT}, such as the
$PT$ form factors of the deuteron \cite{dan} and $\slashP T$ $N\!N$
observables \cite{CPonP}. The results in Refs. \cite{Liu04,Ste08}
suggest that the same is true for EDMs, and we partially confirm this below.
We use the codes from Refs. $\cite{Liu04,Ste08}$, but we recast and
extend the results in the framework of chiral EFT with nonperturbative OPE. 
In particular, we apply power counting in order to make more
model-independent statements. The cases of the helion and the triton
are typical of a generic nucleus. However,
in the deuteron, because of its isoscalar character,
the formally LO contribution from the $\tb$ term 
vanishes \cite{Khrip,Liu04,Vri11b},
a property expected \cite{HaxHenley83} for nuclei with equal number
of protons and neutrons, $N=Z$. We exploit the systematic character
of EFT to extend the deuteron calculation for the $\tb$ term 
to the first non-vanishing order.

Our article is organized as follows. In Section 2, we present
the $P$- and $T$-conserving and violating interactions relevant
for the calculation of light nuclear EDMs.
In Section 3 we discuss in general the power counting of the various
contributions, and present the leading $\slashPT$ potentials and currents,
while in Section 4 we specifically address nuclei with $N=Z$. Next,
we evaluate the EDM of the deuteron in Section 5 and the EDMs of
the helion and the triton in Section 6. In Section 7 we discuss our results
and their implications. Appendices are devoted to the expression of
potential and currents in coordinate space.

\section{Chiral Perturbation Theory}

$\chi$PT is the EFT of QCD for processes involving momenta 
$Q\sim m_{\pi}\ll\MQCD$,
where $m_{\pi}$ is the pion mass and $\MQCD\sim1$ GeV is the characteristic
scale of QCD. At such momenta the relevant degrees of freedom are
nucleons, photons, and pions. The (approximate) chiral symmetry of
QCD, $SU_{L}(2)\times SU_{R}(2)\sim SO(4)$, plays a very important
role, because it constrains the form of the interactions involving
the (pseudo) Goldstone bosons associated with its spontaneous breaking,
the pions. In this section we review these interactions, in both $PT$
and $\slashPT$ sectors of the theory.

The $\chi$PT Lagrangian contains all interactions allowed by the
symmetries of QCD. Each interaction is written in terms of pion ($\boldpi$),
nucleon ($N$), and photon ($A_{\mu}$) fields. The constraints of
(global) chiral and (gauge) electromagnetic symmetries can be incorporated
through the use of covariant derivatives, 
\begin{equation}
\left(D_{\mu}\boldpi\right)_{a}=\frac{1}{D}
\left(\partial_{\mu}\delta_{ab} + eA_{\mu}\varepsilon_{3ab}\right)\pi_{b}
\label{piond}
\end{equation}
for the pion, 
\begin{equation}
{\mathcal{D}}_{\mu}N=
\left[\partial_{\mu}+\frac{i}{F_{\pi}^{2}}
\boldtau\cdot\left(\boldpi\times D_{\mu}\boldpi\right)
+ ieA_{\mu}\frac{1+\tau_{3}}{2}\right]N
\label{nucd}
\end{equation}
for the nucleon, and 
\begin{equation}
F_{\mu\nu}=\partial_{\mu}A_{\nu}-\partial_{\nu}A_{\mu}
\end{equation}
for the photon. Here $F_{\pi}\simeq$ 185 MeV is the pion decay constant,
\begin{equation}
D\equiv1+\frac{\boldpi^{2}}{F_{\pi}^{2}} \ ,
\end{equation}
 $\boldtau$ are the Pauli matrices in isospin space, and $e$ is
the proton electric charge.
Since $m_{N}\sim\MQCD$, nucleons are approximately non-relativistic
in the processes of interest, and Lorentz invariance is incorporated
order by order in the EFT expansion \cite{HBChPT}. We denote the (small)
nucleon four-velocity by $v_{\mu}$ and its spin by $S_{\mu}$; in
the nucleon rest frame, $v_{\mu}=(1,\vec 0)$ and $S^{\mu}=(0,\vec \sigma/2)$
in terms of the Pauli matrices in spin space, $\vec \sigma$.

Chiral-invariant interactions are built with pion covariant derivatives,
while explicit chiral symmetry breaking is introduced 
by the average quark mass $\bar{m}=(m_{u}+m_{d})/2$, 
by the quark mass difference $m_{d}-m_{u}=2\bar{m}\varepsilon$, 
by electromagnetic interactions,
and $\slashP$ and/or $\slashT$ interactions. 
Each interaction with
the correct symmetry transformation properties has a strength determined
by details of the QCD dynamics. Until they are known, they are estimated
using naive dimensional analysis (NDA) \cite{weinberg79,NDA,Weinberg:1989dx}.
For example, the pion mass term originates from explicit chiral-symmetry
breaking by the average quark mass,
so $m_{\pi}^{2}=\mathcal{O}(\bar{m}\MQCD)$. The LECs of other chiral-breaking
interactions proportional to powers of $\bar{m}$ and $\bar{m}\varepsilon$
can then be written in terms of $m_{\pi}^{2}/\MQCD$. Exchange of
hard photons leads to interactions among hadrons that are proportional
to the fine-structure constant $\alpha_{\textrm{em}}=e^{2}/4\pi$.
For simplicity, we count $\varepsilon\sim1/3$ as ${\cal O}(1)$ and
$\alpha_{\textrm{em}}/4\pi$ as ${\cal O}(m_{\pi}^{3}/\MQCD^{3})$,
since numerically 
$\alpha_{\textrm{em}}/4\pi\sim\varepsilon m_{\pi}^{3}/(2\pi F_{\pi})^{3}$.
It is convenient to organize the infinity of effective interactions in the Lagrangian using an
integer {}``chiral index'' $\Delta$ and the number $f$ of fermion
fields \cite{weinberg79,Weinbergbook}: 
\begin{equation}
\mathcal{L}=\sum_{\Delta=0}^{\infty}\sum_{f}\mathcal{L}_{f}^{(\Delta)} \ ,
\label{lagrpc}
\end{equation}
where $\Delta=d+f/2-2\ge0$, with $d$ the number of covariant derivatives
and powers of $m_{\pi}$. The index $\Delta$ tracks the number of
powers of $\MQCD^{-1}$.

\subsection{$P$- and $T$-conserving chiral Lagrangian}

The calculation of the $\slashPT$ potential and currents, which we
need in order to evaluate nuclear EDMs, requires also
$PT$ interactions, which we
summarize here. (A more complete list can be found in, for example,
Refs. \cite{ulfreview,nEFT,Bernard:1992qa,vanKolck,Friar:2004ca}.) 
These interactions result
from the quark (color-gauged) kinetic and mass terms in the QCD Lagrangian.

The terms we need in the $PT$ Lagrangian are 
\begin{eqnarray}
\mathcal{L}_{f\le2,PT}^{(0,1,2)} & = & 
\frac{1}{2}D_{\mu}\boldpi\cdot D^{\mu}\boldpi
-\frac{m_{\pi}^{2}}{2D}\boldpi^{2}
+\bar{N}iv\cdot\mathcal{D}N
-\frac{1}{2m_{N}}\bar{N}\left[\mathcal{D}^{2}-(v\cdot\mathcal{D})^{2}\right]N 
\nonumber \\
& & -\frac{\breve{\delta}m_{\pi}^{2}
-\delta m_{N}^{2}}{2D^{2}}\left(\boldpi^{2}-\pi_{3}^{2}\right)
-\frac{\delta^{}m_{\pi}^{2}}{2D^{2}}\pi_{3}^{2}
-\left(\delta m_{N}+\breve{\delta}m_{N}\right)
\left(\boldpi\times v\cdot D\,\boldpi\right)_{3} 
\nonumber \\
& & -\frac{2g_{A}}{F_{\pi}}D_{\mu}\boldpi\cdot\bar{N}\boldtau S^{\mu}N
+\frac{\beta_{1}}{F_{\pi}}\left(D_{\mu}\pi_{3}-\frac{2\pi_{3}}{F_{\pi}^{2}D}
\boldpi\cdot D_{\mu}\boldpi\right)\bar{N}S^{\mu}N
\nonumber \\
& & -\frac{g_{A}\delta m_{N}}{F_{\pi}m_{N}}\left[i\bar{N}
\left(\boldtau\times\boldpi\right)_{3}S\cdot\mathcal{D} N+
\mathrm{H.c.}\right] 
\nonumber \\
& & -\frac{e}{16m_{N}^{2}}\varepsilon^{\alpha\beta\mu\nu} F_{\mu\nu}
\left\{ i\bar{N}\left[1+2\kappa_{0}+\left(1+2\kappa_{1}\right)\tau_{3}\right]
S_{\alpha}\mathcal{D}_{\beta}N+\mathrm{H.c.}\right\} \ .
\label{LagStrong012}
\end{eqnarray}
The pion kinetic and mass terms have $\Delta=0$. For notational
simplicity, we choose to absorb in the pion mass $m_{\pi}$ a correction
$\propto{\bar{m}}^{2}$. At $\Delta=1$, the leading electromagnetic
contribution to the pion mass splitting appears, 
$\breve{\delta}m_{\pi}^{2}=\mathcal{O}(\alpha_{\textrm{em}}\MQCD^{2}/4\pi)$,
while the quark-mass difference contribution, $\delta m_{\pi}^{2}=
\mathcal{O}(\varepsilon^{2}m_{\pi}^{4}/\MQCD^{2})$,
is smaller by one power of $\varepsilon m_{\pi}/\MQCD$. The pion
mass splitting, $m_{\pi^{\pm}}^{2}-m_{\pi^{0}}^{2}=\breve{\delta}m_{\pi}^{2}
-\delta^{}m_{\pi}^{2}=(35.5\;{\rm MeV})^{2}$
\cite{Nakamura:2010zzi}, is dominated by the electromagnetic contribution.
The nucleon kinetic terms have $\Delta=0,1$. Again for simplicity,
the average nucleon mass $m_{N}$ absorbs a correction $\propto\bar{m}$,
the sigma term. Entering at $\Delta=1,2$, the nucleon mass splitting,
$m_{n}-m_{p}=\delta m_{N}^{}+\breve{\delta}m_{N}^{}=1.29$ MeV 
\cite{Nakamura:2010zzi}
also receives contributions from electromagnetism and from the quark
masses. In this case, the quark-mass contribution $\delta m_{N}$
is expected to be the largest. By dimensional analysis
$\delta m_{N}={\cal O}(\varepsilon m_{\pi}^{2}/\MQCD)$,
and lattice simulations estimate it to be 
$\delta m_{N}=2.26\pm0.57\pm0.42\pm0.10$
MeV \cite{latticedeltamN}, which is in agreement with an extraction
from charge-symmetry breaking in the $pn\to d\pi^{0}$ reaction \cite{CSBd}.
The electromagnetic contribution is $\breve{\delta}m_{N}=
\mathcal{O}(\alpha_{{\rm {em}}}\MQCD/4\pi)$,
which is $\mathcal{O}(\varepsilon m_{\pi}^{3}/\MQCD^{2})$ and about
the $20\%$ of $\delta m_{N}$. From the Cottingham sum rule \cite{Cott}
one finds $\breve{\delta}m_{N}=-(0.76\pm0.30)$ MeV, which is consistent
with dimensional analysis. In order to achieve the form \eqref{LagStrong012}
for $\mathcal{L}_{PT}$, we have used a field redefinition \cite{Friar:2004ca}
to eliminate the nucleon mass difference term in favor of pionic mass
and interaction terms. In this way, the nucleon mass to be used in
nucleon propagators
%the solution of the Schr\"odinger or Lippmann-Schwinger equations 
is simply $m_{N}$. The operator with LEC $g_{A}$ is the usual pion-nucleon
axial coupling appearing at $\Delta=0$. We also absorb subleading
corrections in it, so that the Goldberger-Treiman relation for the
strong pion-nucleon constant, $g_{N\!N\pi}=2g_{A}\,m_{N}/F_{\pi}$, applies
without an explicit discrepancy. If for the pion-nucleon coupling
constant we use $g_{N\!N\pi}=13.07$ \cite{piNcoupling}, then $g_{A}=1.29$.
Its isospin-violating counterpart is the operator with LEC
$\beta_{1}=\mathcal{O}(\varepsilon m_{\pi}^{2}/\MQCD^{2})$
at $\Delta=2$. At present there are only bounds on $\beta_{1}$ from
isospin violation in $N\! N$ scattering. The Nijmegen partial-wave
analysis of $N\! N$ scattering data gives $\beta_{1}=(0\pm9)\cdot10^{-3}$
\cite{isoviolphen}, which is comparable to estimates of $\beta_{1}$
from $\pi$-$\eta$ mixing. 
At $\Delta=2$ there is another isospin-violating
pion-nucleon interaction generated by nucleon recoil and the nucleon
mass difference. 
Finally, also at $\Delta=2$ there is a relativistic
correction \cite{Manohar} to the electromagnetic coupling of the nucleon
involving
the isoscalar and isovector components of the anomalous magnetic moment,
respectively $\kappa_{0}=-0.12$ and $\kappa_{1}=3.7$.

\subsection{$P$- and $T$-violating chiral Lagrangian}

\label{ToddLag}

The lowest-dimension $\slashPT$ operator that can be added to the
$PT$ QCD Lagrangian is the dimension-four $\bar{\theta}$ term. With
an appropriate choice of the quark fields $q=(u,d)^{T}$, the $\bar{\theta}$
term can be expressed as a complex mass term \cite{Bal79}, 
\begin{equation}
\mathcal{L}_{\slashPTsub, {\rm dim=4}} 
= m_{\star}\bar{\theta}\;\bar{q}i\gamma_{5}q\ ,
\label{LtrvQCD}
\end{equation}
where $m_{\star}=m_{u}m_{d}/(m_{u}+m_{d})=\mathcal{O}(m_{\pi}^{2}/\MQCD)$
and $\bar{\theta}$ is the QCD vacuum angle, here already assumed
to be small, $\bar{\theta}\lesssim10^{-10}$, as indicated by the
experimental bound on the neutron EDM.

The smallness of $\bar{\theta}$ leaves room for other $\slashPT$
sources in the strong interactions, which have their origin in an
ultraviolet-complete theory at a high energy scale, such as, for example,
supersymmetric extensions of the Standard Model \cite{RamseyMusolf:2006vr}.
Well below the scale $M_{\slashT}$ characteristic of $T$ violation,
we expect $\slashPT$ effects to be captured by the lowest-dimension
interactions among Standard Model fields that respect 
$SU_{c}(3)\times SU_{L}(2)\times U_{Y}(1)$
gauge symmetry. Above $\MQCD$, strong interactions are described
by the most general Lagrangian with Lorentz, color, and electromagnetic
gauge invariance among the lightest quarks, gluons, and photons. The
effectively dimension-six $\slashPT$ terms at this scale can be written
as \cite{dim6origin,Weinberg:1989dx,Grzadkowski:2010es,RamseyMusolf:2006vr}
\begin{eqnarray}
{\cal L}_{\slashPTsub,{\rm dim=6}} & = & 
-\frac{1}{2}\bar{q}\left(d_{0}+d_{3}\tau_{3}\right)\sigma^{\mu\nu}
i\gamma^{5}q\; F_{\mu\nu}
-\frac{1}{2}\bar{q}\left(\tilde{d}_{0}+\tilde{d}_{3}\tau_{3}\right)
\sigma^{\mu\nu}i\gamma^{5}\lambda^{a}q\; G_{\mu\nu}^{a}
\nonumber \\
& & +\frac{d_{W}}{6}\varepsilon^{\mu\nu\lambda\sigma}f^{abc}
G_{\mu\rho}^{a}G_{\nu}^{b\,\rho}G_{\lambda\sigma}^{c}
+\frac{1}{4}\textrm{Im}{\Sigma_{1}}\left(\bar{q}q\,\bar{q}i\gamma^{5}q
-\bar{q}\,\boldtau q\,\cdot\bar{q}\,\boldtau i\gamma^{5}q\right)
\nonumber \\
& & +\frac{1}{4}\textrm{Im}{\Sigma_{8}}\left(\bar{q}\lambda^{a}q\,
\bar{q}\lambda^{a}i\gamma^{5}q
-\bar{q}\lambda^{a}\,\boldtau q\,\cdot\bar{q}\lambda^{a}\,\boldtau
i\gamma^{5}q\right)\ ,
\label{eq:dim6}
\end{eqnarray}
in terms of the 
gluon field strength
$G_{\mu\nu}^{a}$, 
the Gell-Mann matrices $\lambda^{a}$ in color space, and the associated
structure constants $f^{abc}$. In Eq. (\ref{eq:dim6}) the first
(second) term represents the isoscalar $d_{0}$ ($\tilde{d}_{0}$)
and isovector $d_{3}$ ($\tilde{d}_{3}$) components of the qEDM (qCEDM).
Although these interactions have canonical dimension five, they originate
just above the Standard Model scale $M_{W}$ from dimension-six operators
\cite{dim6origin} involving in addition the carrier of electroweak
symmetry breaking, the Higgs field. They are thus proportional to
the vacuum expectation value of the Higgs field, which can be traded
in for the ratio of the quark mass to Yukawa coupling, $m_{q}/f_{q}$.
Writing the proportionality constant as $e\delta_{q}f_{q}/M_{\slashT}^{2}$
($4\pi\tilde{\delta}_{q}f_{q}/M_{\slashT}^{2}$), we have
\begin{equation}
d_{0,3}\sim\mathcal{O}\left(e\delta_{0,3}
\frac{\bar{m}}{M_{\slashT}^{2}}\right)\ ,
\qquad\tilde{d}_{0,3}\sim \mathcal{O}\left(4\pi\tilde{\delta}_{0,3}
\frac{\bar{m}}{M_{\slashT}^{2}}\right)\ ,
\end{equation}
in terms of the average light-quark mass $\bar{m}$ and the dimensionless
factors $\delta_{0,3}$ and $\tilde{\delta}_{0,3}$ that represent
typical values of $\delta_{q}$ and $\tilde{\delta}_{q}$. The third
term in Eq. (\ref{eq:dim6}) \cite{Weinberg:1989dx} is the gCEDM,
with coefficient 
\begin{equation}
d_{W}\sim\mathcal{O}\left(\frac{4\pi w}{M_{\slashT}^{2}}\right)\ ,
\label{w}
\end{equation}
in terms of a dimensionless parameter $w$. The fourth and fifth
operators \cite{Grzadkowski:2010es,RamseyMusolf:2006vr} are $\slashPT$
FQ operators, with coefficients 
\begin{equation}
\textrm{Im}\Sigma_{1,8}=
\mathcal{O}\left(\frac{(4\pi)^{2}\sigma_{1,8}}{M_{\slashT}^{2}}\right)\ ,
\label{sigma}
\end{equation}
in terms of further dimensionless parameters $\sigma_{1,8}$. The
sizes of $\delta_{0,3}$, $\tilde{\delta}_{0,3}$, $w$, and $\sigma_{1,8}$
depend on the exact mechanisms of electroweak and $PT$ breaking and
on the running to low energies where nonperturbative QCD sets in.
The minimal assumption is that they are $\mathcal{O}(1)$, 
$\mathcal{O}(g_{s}/4\pi)$,
$\mathcal{O}((g_{s}/4\pi)^{3})$, and $\mathcal{O}(1)$, respectively,
with $g_{s}$ the strong coupling constant. However, they could be
significantly smaller, when parameters encoding $\,\slashPT$ beyond
the Standard Model are small, or significantly larger, since $f_{q}$
is unnaturally small; for discussion and examples, see for instance
Refs. \cite{Pospelov:2005pr,RamseyMusolf:2006vr}.

The operators in Eqs. (\ref{LtrvQCD}) and (\ref{eq:dim6}) 
have different transformation properties under chiral symmetry, which
has profound implications for the form and relative importance of the
$\slashPT$ pion-nucleon and $N\! N$ couplings in the effective Lagrangian.
The $\bar{\theta}$ term in Eq. (\ref{LtrvQCD}) transforms 
as the fourth component of an $SO(4)$ vector 
$P=(\bar{q}\,\boldtau q,\bar{q}i\gamma_{5}q)$,
the third component of which is responsible for quark-mass isospin
violation \cite{vanKolck}. $\slashPT$ from the $\bar{\theta}$ term
and isospin violation from the quark mass difference are therefore
intrinsically linked; this link appears in certain relations 
\cite{Bal79,Cre79,BiraEmanuele}
between the coefficients of $\slashPT$ and isospin-breaking operators
in $\chi$PT through a coefficient 
$\rho=(1-\varepsilon^{2})\bar{\theta}/2\varepsilon$.
The dimension-six operators in Eq. (\ref{eq:dim6}) have different
transformation properties still \cite{dim6,Vri11a,Mer11,Vri11b,TVpotential}.
The isoscalar and isovector qEDM and qCEDM transform as the fourth
and third components of two other $SO(4)$ vectors. 
There is no useful link to $PT$ observables, and the third component
of the qCEDM vector tends to generate hadronic interactions, which
for $\bar{\theta}$ require tensor products and are of higher order.
For qEDM, purely hadronic interactions arise from integrating out at least
one hard photon, which leads to further breaking of chiral symmetry
in the form of tensor products of the vectors 
with an antisymmetric chiral tensor \cite{vanKolck}. The contributions
of the qEDM to 
%purely 
hadronic couplings, like pion-nucleon or 
$N\! N$ couplings, are suppressed by $\alpha_{\textrm{em}}/4\pi$. In contrast,
the gCEDM and the two $\slashPT$ FQ operators
are singlets of the chiral group. Because they are chiral invariant,
and contain no photons, 
the gCEDM and the two $\slashPT$ FQ operators
lead to exactly the same effective interactions, although, of course,
with different strengths. 
For simplicity of notation, in the following we treat gCEDM and $\slashPT$
FQ operators together; we refer to them as chiral-invariant ($\chi$I) sources
and use $w$ to denote both $w$ and $\sigma_{1,8}$: 
\begin{equation}
\{w,\sigma_{1},\sigma_{8}\}\to w.
\end{equation}

We now present a subset of the complete $\slashPT$ chiral Lagrangian
originating from the fundamental sources above. We only give the operators
that play a role in the LO calculation of light-nuclei EDMs, the more
general Lagrangian being found in Refs. \cite{BiraEmanuele,dim6}.
In general, a LO calculation of the EDM of a light nucleus
requires \textit{six} $\slashPT$ interactions:
\begin{eqnarray}
\mathcal{L}_{\slashPTsub} & = & 
- 2\, \Nb\left(\bar{d}_{0}+\bar{d}_{1}\tau_{3}\right)S^{\mu}N\, v^{\nu}\Fmu
-\frac{1}{F_{\pi}}
\bar{N}\left(\bar{g}_{0}\,\boldtau\cdot\boldpi+\bar{g}_{1}\pi_{3}\right)N
\nonumber \\
& & +\bar{C}_{1}\bar{N}N \, \partial_{\mu}(\bar{N}S^{\mu}N)
+\bar{C}_{2}\bar{N}\boldtau N\cdot
\partial_{\mu}(\bar{N}S^{\mu}\boldtau N)
+\ldots\ ,
\label{chiLag123}
\end{eqnarray}
which represent short-range isoscalar ($\bar{d}_{0}$) and isovector
($\bar{d}_{1}$) contributions to the nucleon EDM, isoscalar ($\bar{g}_{0}$)
and isovector ($\bar{g}_{1}$) non-derivative pion-nucleon couplings,
and two short-range $\slashPT$ $N\! N$ interactions ($\bar{C}_{1}$,
$\bar{C}_{2}$). 
Here we relegate to the {}``$\ldots$'' terms related to the above
by chiral symmetry. The explicit forms of these terms depend on the
$\slashPT$ source but, because they involve more pion fields, they
do not appear in the LO EDMs we are interested in. 
Note that Eq. \eqref{chiLag123}
is the form of $\mathcal{L}_{\slashPTsub}$ after a field redefinition
is performed to eliminate pion tadpoles and guarantee vacuum alignment;
the parameters thus absorb contributions generated by this field redefinition.

Which of these six interactions is relevant depends on the system
we are studying and on the fundamental $\slashPT$ source. As will
be seen, the spin and isospin of the deuteron cause the deuteron EDM
to be sensitive to only three of the above operators. In more general
cases, the EDMs of light nuclei are sensitive to all six interactions.
The EDMs of heavy nuclei could involve more operators than the set
above.
Generically one might expect a dominance by effects from ($i$)
a single nucleon, since multi-nucleon contributions tend to be suppressed
at low energies by phase space; and ($ii$) pions, thanks to
their small mass and related long range. However, significant deviation
from this expectation comes from the relative sizes of the various
LECs, which depends on the $\slashPT$ source. NDA leads to the following
estimates for the dimension-four and -six $\slashPT$ sources:

\begin{itemize}

\item For the $\tb$ term, four operators play a role at LO, the other two
appearing only at subleading orders. In order to generate $\bar{g}_{1}$,
which is relevant for the deuteron EDM, the $\tb$ term requires an
insertion of the quark mass difference, which causes a suppression
of $\bar{g}_{1}$ relative to $\bar{g}_{0}$ by a factor 
$\varepsilon\mpi^{2}/\MQCD^{2}$
\cite{BiraEmanuele}. (At the same order, there exists also a two-derivative
pion-nucleon coupling, but for our purpose here it can be absorbed
by a small change in $\bar{g}_{0}$ \cite{TVpotential}.) The LECs
scale as 
\begin{equation}
\bar{g}_{0}=\Or\!\left(\bar{\theta}\frac{m_{\pi}^{2}}{\MQCD}\right),
\qquad
\bar{g}_{1}=\Or\!\left(\varepsilon\bar{\theta}\frac{m_{\pi}^{4}}{\MQCD^{3}}
\right),
\qquad
\bar{d}_{0,1}=\Or\!\left(e\bar{\theta}\frac{m_{\pi}^{2}}{\MQCD^{3}}\right).
\label{NDAtheta}
\end{equation}

\item For the qCEDM, the same four operators are needed. 
In this case, there is no \textit{a priori} relative suppression of
$\bar{g}_{1}$ and the LECs scale as 
\begin{eqnarray}
\bar{g}_{0}&=&\Or\!\left((\tilde{\delta}_{0}+\varepsilon\tilde{\delta}_{3})
\frac{m_{\pi}^{2}\MQCD}{M_{\slashT}^{2}}\right),\qquad
\bar{g}_{1}=\Or\!\left(\tilde{\delta}_{3}
\frac{m_{\pi}^{2}\MQCD}{M_{\slashT}^{2}}\right),\; \nonumber\\
\bar{d}_{0,1} &=&\Or\!\left((\tilde{\delta}_{0} + \tilde{\delta}_{3})
\frac{m_{\pi}^{2}}{\MQCD M_{\slashT}^{2}}\right).
\label{NDAqCEDM}
\end{eqnarray}
(Here the ``+'' signs are not to be taken literally; they are only
meant to signify two independent contributions to a LEC.)

\item For the qEDM, only the short-range EDM contributions are important,
and they scale as 
\begin{equation}
\bar{d}_{0,1}=\Or\left(e\delta_{0,3}
\frac{m_{\pi}^{2}}{\MQCD M_{\slashT}^{2}}\right).
\label{NDAqEDM}
\end{equation}

\item For the $\chi$I (gCEDM and FQ) $\slashPT$ sources, the non-derivative
pion-nucleon interactions, which break chiral symmetry, are suppressed
by a factor $\mpi^{2}/\MQCD^{2}$ compared to short-range nucleon
EDM contributions and $\slashPT$ $N\! N$ interactions, which conserve
chiral symmetry. (Again, a two-derivative pion-nucleon interaction
exists at the same order but can be absorbed in $\bar{g}_{0}$ 
\cite{TVpotential}.)
All six operators thus become relevant, and the LECs scale as 
\begin{eqnarray}
\bar{g}_{0}=\Or\!\left(w\frac{m_{\pi}^{2}\MQCD}{M_{\slashT}^{2}}\right), 
& & \bar{g}_{1}=\Or\!\left(\varepsilon w
\frac{m_{\pi}^{2}\MQCD}{M_{\slashT}^{2}}\right),
\nonumber \\
\bar{d}_{0,1}=\Or\!\left(ew\frac{\MQCD}{M_{\slashT}^{2}}\right), 
& & \bar{C}_{1,2}=\Or\left(w\frac{\MQCD}{\Fp^{2}M_{\slashT}^{2}}\right).
\label{NDATVCI}
\end{eqnarray}

\end{itemize}

\subsection{EDM of the nucleon}
\label{NucEDM}

Using these interactions, the nucleon EDM has been calculated in $\chi$PT
up to NLO for all sources of dimension up to six \cite{Ottnad,Vri11a,Mer11}. 

In the power counting of $\chi$PT \cite{weinberg79},
one considers typical momenta $Q\sim\mpi\sim\Fp\ll\MQCD\sim m_{N}\sim2\pi\Fp$
and assigns
\begin{itemize}
\item a factor $Q^{4}/(4\pi)^{2}$ for each loop integral; 
\item a factor $1/Q$ for each nucleon propagator; 
\item a factor $1/Q^{2}$ for each pion propagator; 
\item the NDA estimate for the LECs corresponding to the interactions in
the diagram. 
\end{itemize}
This produces for any observable an expansion in the small ratio $Q/\MQCD$.

For the nucleon EDM,
in all cases there are short-range contributions from $\bar{d}_{0,1}$ at LO. 
For qEDM and $\chi$I sources, the relative suppression of pion-nucleon
couplings 
means that loops come at higher orders
and only $\bar{d}_{0,1}$ appear up to NNLO \cite{Vri11a}. 
In contrast, for $\bar{\theta}$ and qCEDM, one-loop diagrams contribute
at LO and NLO. 
Using dimensional
regularization in $d$ dimensions at a renormalization scale $\mu$,
and introducing 
\begin{equation}
\delta\bar{d}_{1}\equiv\frac{eg_{A}\bar{g}_{0}}{(2\pi F_{\pi})^{2}}
\left(\frac{2}{4-d}-\gamma_{E}+\ln\frac{4\pi\mu^{2}}{m_{N}^{2}}\right)\,
\end{equation}
where $\gamma_{E}\simeq 0.577$ is the Euler-Mascheroni constant, the
isoscalar and isovector EDMs can be expressed at NLO respectively
as \cite{Ottnad,Vri11a,Mer11}
\begin{equation}
d_{0}=\bar{d}_{0}
+\frac{eg_{A}\bar{g}_{0}}{(2\pi F_{\pi})^{2}}
\pi \left[\frac{3m_\pi}{4m_N}-\frac{\delta m_N}{m_\pi}\right]
+\frac{eg_{A}\bar g_1}{(2\pi F_{\pi})^{2}}\frac{\pi}{4}\frac{m_\pi}{m_N}
\label{d0}
\end{equation}
and
\begin{equation}
d_{1}=\bar{d}_{1}+\delta\bar{d}_{1}
+\frac{eg_{A}\bar{g}_{0}}{(2\pi F_{\pi})^{2}}
\left[\ln\frac{m_{N}^{2}}{m_{\pi}^{2}}+\frac{5\pi}{4}\frac{m_\pi}{m_N}
-\frac{\breve\delta m_\pi^2}{m_\pi^2}
\right]
+\frac{eg_{A}\bar g_1}{(2\pi F_{\pi})^{2}}\frac{\pi}{4}\frac{m_\pi}{m_N},
\label{d1}
\end{equation}
where the $\bar g_1$ terms applies to qCEDM only. 
The dependence on the arbitrary scale $\mu$ in $\delta\bar{d}_{1}$
is compensated by $\bar{d}_{1}$.
In fact, the loop contributions cannot be separated from the short-range
pieces in a model-independent way.
After absorbing all these terms in $\bar d_{0,1}$, which we do for
the rest of the paper, we can write for all sources
 \begin{equation}
d_{n}=\bar{d}_{0}-\bar{d}_{1}
\label{dnLO}
\end{equation}
for the neutron and
\begin{equation}
d_{p}=\bar{d}_{0}+\bar{d}_{1}
\label{dpLO}
\end{equation}
for the proton.

However, one expects no cancellation between short-range contributions, 
which are analytic in $m_{\pi}^{2}$, 
and the {}``chiral-log'' and other finite terms, which are not. 
Thus the non-analytic terms serve as lower-bound
estimates for 
the size of $d_{p,n}$. 
We then expect, for $\bar{\theta}$ \cite{Cre79}
and qCEDM \cite{Mer11},
\begin{eqnarray}
\bar d_0 &\gtrsim& 
0.01 
\left[\frac{\bar g_0}{F_{\pi}}+ 0.3 \frac{\bar g_1}{F_{\pi}}\right] 
e\,\textrm{fm} \ ,
\label{d0est}
\\
\bar d_1 &\sim&  
0.1 
\left[\frac{\bar g_0}{F_{\pi}}+ 0.03 \frac{\bar g_1}{F_{\pi}}\right]
e\,\textrm{fm} \ .
\label{d1est}
\end{eqnarray}

As stressed in Ref. \cite{Vri11a}, measurements of both $d_{n}$
and $d_{p}$ alone can tell us little about the underlying source
of $\slashPT$. More can be learned from measuring the EDMs of light
nuclei, the calculation of which we now turn to.

\section{Ingredients: the generic case}
\label{generic}

The EDM of a nucleus with $A\ge2$ nucleons can be separated into
two contributions. The first contribution comes from an insertion
of the $\slashPT$ electromagnetic current $J_{\slashPTsub}^{0}$. The
second stems from the $PT$ charge density $J_{PT}^{0}$ upon perturbing
the wavefunction of the nucleus with the $\slashPT$ potential $V_{\slashPTsub}$,
such that the wavefunction obtains a $\slashPT$ component. 
To first order in the $\slashPT$ sources, the EDM
is thus a sum of two reduced matrix elements 
\begin{equation}
d_{A}=\frac{1}{\sqrt{6}}\left(\left\langle \Psi_{A}
\left|\left|\vec{D}_{\slashPTsub}\right|\right|\Psi_{A}\right\rangle 
+2\,\left\langle \Psi_{A}\left|\left|\vec{D}_{PT}\right|\right|
\widetilde{\Psi}_{A}\right\rangle \right)\,.
\end{equation}
The nuclear ground state $|\Psi_{A}\rangle$ and its parity admixture
$|\widetilde{\Psi}_{A}\rangle$ are the solutions of homogeneous and
inhomogeneous Schr\"odinger equations, 
\begin{eqnarray}
(E-H_{PT})|\Psi_{A}\rangle & = & 0\ ,\\
(E-H_{PT})|\widetilde{\Psi}_{A}\rangle & = & V_{\slashPTsub}|\Psi_{A}\rangle\ ,
\end{eqnarray}
respectively, where $H_{PT}$ is the $PT$ Hamiltonian. 
The $\slashPT$ potential $V_{\slashPTsub}$ is shown in coordinate
space in Appendix \ref{Vcoord}.
The EDM operators
$\vec{D}_{PT}$ and $\vec{D}_{\slashPTsub}$ are obtained from the corresponding
charge densities $J_{PT}^{0}$ and $J_{\slashPTsub}^{0}$, respectively,
as discussed in Appendix \ref{Fourier}. The factor of 2 in front
of the second matrix element corresponds to the number of time-ordered
diagrams, and the phases of wavefunctions are chosen so that these
matrix elements are purely real.

In this section we identify the ingredients needed for the LO calculation
of $d_{A}$, assuming no particular cancellations or suppressions
due to spin/isospin factors.

\subsection{Power counting}

\begin{figure}[t]
\centering \includegraphics[scale = 0.6]{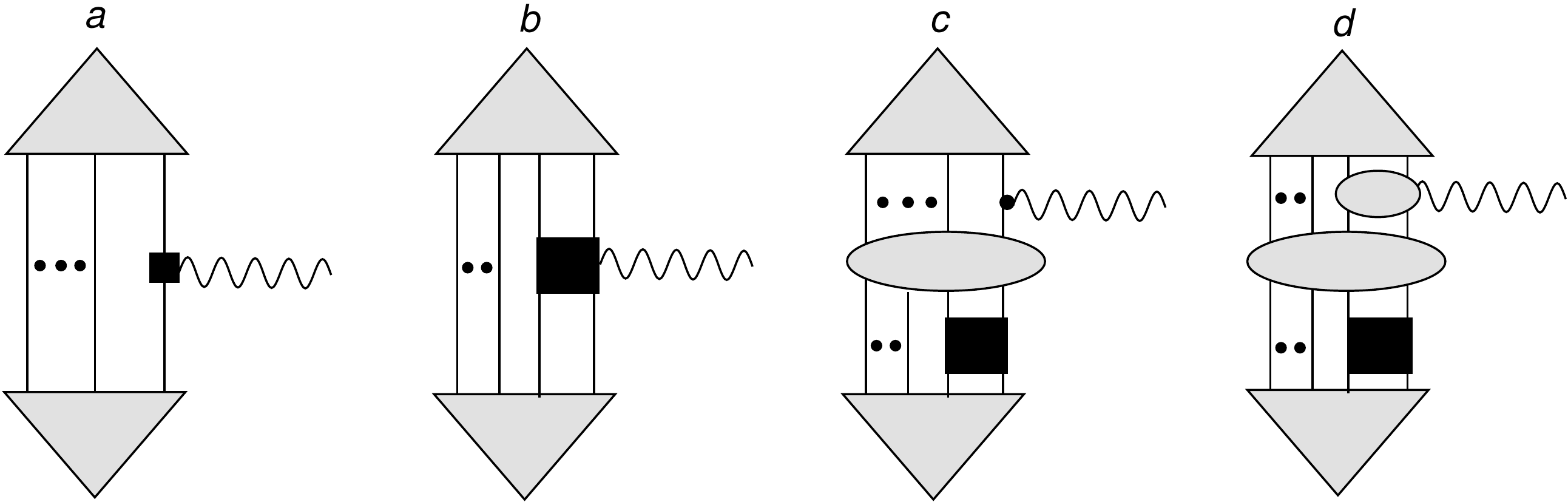}
\caption{The four general classes of diagrams contributing to a nuclear EDM
described in the text. Solid and wavy lines represent nucleons and photons. 
The three (two) dots stand for $A-3$ ($A-4$) nucleon propagators.
The large triangle denotes the nuclear wavefunction;
the oval, iterations of the $PT$ potential; 
the dot with an attached photon, the $PT$ one-body current;
the oval with an attached photon, the $PT$ two-body current; 
the black square, the $\slashPT$ potential;
and the black square with an attached photon, the $\slashPT$ current.}
\label{classes} 
\end{figure}

Both the potential $V_{\slashPTsub}$ and the current $J_{\slashPTsub}^{0}$
can be obtained from the Lagrangian of the previous section. The potential
$V_{\slashPTsub}$ for the various $\slashPT$ sources has been derived
in Ref. \cite{TVpotential}. To the order we are concerned with here,
the potential can be taken as two-body. The $\slashPT$ and $PT$ currents 
can also be divided into one-body and more-body currents. As we will
see, the latter are dominated by two-body effects as well. There are
thus four classes of contributions to a nuclear EDM, schematically
drawn in Fig. \ref{classes}. 
In order to determine which diagram(s) give(s) the most important
contribution(s) 
we need to estimate their sizes by applying power counting.

%%%%%%%%%%%%%%%%%%%%%
%
%
%%%%%%%%%%%%%%%%%%%%%

We need to count powers of the generic momentum $Q$ in the process,
in order to get an expansion in $Q/\MQCD$. Here $Q$ is given by
the nuclear binding momentum, which for a typical nucleus can be taken
as $Q\sim\mpi\sim\Fp$,
as standard in $\chi$PT. 
However, as pointed out by Weinberg \cite{weinberg}, the power counting
of $\chi$PT needs to be adapted to the existence
for $A\ge2$ of intermediate states consisting purely of propagating nucleons. 
A generic diagram can be split into {}``reducible'' parts, that
contain such states, and {}``irreducible'' subdiagrams, which do
not. Within an irreducible subloop, the contour integration over the
0th component of the loop momentum can always be performed in such
a way as to avoid the nucleon pole. In these diagrams the nucleon
energy is of order $Q$, as assumed in $\chi$PT power counting. 
On the other hand, in diagrams where the intermediate state consists
purely of propagating nucleons, \textit{i.e.} reducible diagrams,
one cannot avoid the poles of nucleon propagators, thus picking up
energies $\sim Q^{2}/m_{N}$ \cite{weinberg} rather than $\sim Q$.
Moreover, such loops also obtain an additional enhancement of $4\pi$.
The contribution of such a reducible diagram can be counted by applying
the modified rules \cite{nEFT}: 
\begin{itemize}
\item a factor $Q^{5}/(4\pi m_{N})$ for each loop integral; 
\item a factor $m_{N}/Q^{2}$ for each nucleon propagator; 
\item a factor $1/Q^{2}$ for each pion propagator; 
\item the NDA estimate for the LECs corresponding to the interactions in
the diagram. 
\end{itemize}
As an example, consider an insertion of a LO, $PT$ pion exchange
in a diagram. It gives rise to one additional loop $\sim Q^{5}/(4\pi m_{N})$,
two nucleon propagators $\sim m_{N}^{2}/Q^{4}$, a pion propagator
$\sim1/Q^{2}$, and two insertions of the strong pion-nucleon vertex
$\sim(Q/\Fp)^{2}$. Combining these factors, the extra one-pion exchange
amounts to $Q/M_{N\! N}$, where $M_{N\! N}=4\pi\Fp^{2}/m_{N}\sim\Fp$.
A similar power counting holds for short-range $PT$ interactions,
although the situation for them is more complicated \cite{morethanweinberg}.
For very light nuclei, $Q<M_{N\! N}$ and pion exchange can be treated
perturbatively \cite{KSW,nEFT}. 
The deuteron EDM has in fact already been considered
in this light \cite{Vri11a}. For less dilute nuclei,
however, one expects $Q\sim M_{N\! N}$ and pion exchange needs to
be summed to all orders \cite{nEFT,morethanweinberg}. The counting rules above are a generalization for $A\ge 2$ of the rules given in Ref. \cite{nEFT}. Note that they provide a natural explanation for the $Q/\MQCD$ supression associated with an additional nucleon observed in pion-nucleus scattering \cite{Liebig:2010ki, juelich}.

We can now estimate the size of each of the classes of diagrams in
Fig. \ref{classes}. 
For each class we take the $PT$ and $\slashPT$ LO interactions in
Eqs. \eqref{LagStrong012} and \eqref{chiLag123}, respectively. The
iteration of the LO $PT$ potential costs no factors, and is necessary
among nucleons in reducible intermediate states, as indicated in diagrams
(c) and (d) of Fig. \ref{classes}. Such iteration among nucleons
before and after all $\slashPT$ and electromagnetic insertions builds up 
the $PT$ wavefunction,
represented in Fig. \ref{classes} as well, which introduces an overall
normalization of the diagrams. This normalization can be read off
from the diagram analogous to (a), where the one-body current is given
instead by the electromagnetic charge. In the following we account
for this normalization by omitting the $A-1$ loops and $A+1$ nucleon
propagators that are common to all diagrams. 
Thus, diagram (a) is simply 
\begin{equation}
D_{a}={\cal O}\left(d_{p,n}Q\right).
\end{equation}
In contrast, diagram (b) has one additional irreducible loop 
$\sim Q^{5}/(4\pi m_{N})$,
one additional nucleon propagator $\sim m_{N}/Q^{2}$, and the leading
$\slashPT$ two-body current. For both qEDM and $\chi$I sources the latter
brings a suppression of a factor $Q^{2}/\MQCD^{2}$, whereas for the
other sources the contribution is comparable to the one-body term.
One can continue in this fashion to find that for diagram (c), 
\begin{equation}
D_{c}={\cal O}\left(e\,\frac{\bar{g}_{0,1}}{F_{\pi}^{2}}Q\right)
+{\cal O}\left(e\,\bar{C}_{1,2}F_{\pi}^{2}Q\right),
\end{equation}
while for diagram (d) there is always a further suppression by a factor 
$Q^{2}/\MQCD^{2}$.
Analogously, more-body potentials and currents bring further suppression.

Plugging in the scaling of the LECs for the different sources, Eqs.
\eqref{NDAtheta}, \eqref{NDAqCEDM}, \eqref{NDAqEDM}, and \eqref{NDATVCI},
we can draw the following general 
expectations for the EDMs of light nuclei: 
\begin{itemize}
\item For the $\tb$ term, the nuclear EDM is dominated by diagram (c):
the nuclear wavefunction acquires a $\slashPT$ admixture after a
one-pion exchange involving the isoscalar $\bar{g}_{0}$ vertex; the
admixed wavefunction then couples to the proton charge. 
\item For the qCEDM, the nuclear EDM is dominated by the same effect as
the $\tb$ term. However, for the qCEDM the $\slashPT$ pion-nucleon
vertex can be either $\bar{g}_{0}$ or $\bar{g}_{1}$. 
\item For the qEDM, the nuclear EDM is dominated by the sum of the EDMs
of the constituent nucleons, diagram (a). 
\item For $\chi$I 
sources, the nuclear EDM is more complicated than for the other sources.
Due to the chiral suppression of the pion-nucleon interactions, diagrams
(a) and (c) are equally important, and in the latter the short-range
$\slashPT$ interactions $\bar{C}_{1,2}$ need to be included besides
the one-pion exchange from both $\bar{g}_{0}$ and $\bar{g}_{1}$
couplings. 
\end{itemize}

\subsection{$P$- and $T$-odd potential}
\label{LOpotmom}

For all sources considered, except qEDM, an insertion of the LO $\slashPT$
two-nucleon potential appears in the EDM at LO. The general $\slashPT$
$N\! N$ potential was derived in Ref. \cite{TVpotential} and we
summarize the relevant parts here. In momentum space the potential
is given by 
\begin{eqnarray}
V_{\slashPTsub}(\vec{k}\,) & = & 
i\frac{g_{A}\bar{g}_{0}}{F_{\pi}^{2}}\,\boldtau^{(i)}\cdot\boldtau^{(j)}
\left(\vec{\sigma}^{\,(i)}-\vec{\sigma}^{\,(j)}\right)\cdot
\frac{\vec{k}}{\vec{k}^{\,2}+m_{\pi}^{2}}
\nonumber \\ 
& & +i\frac{g_{A}\bar{g}_{1}}{2F_{\pi}^{2}}
\left[\left(\tau_{3}^{(i)}+\tau_{3}^{(j)}\right)
\left(\vec{\sigma}^{\,(i)}-\vec{\sigma}^{\,(j)}\right)
+\left(\tau_{3}^{(i)}-\tau_{3}^{(j)}\right)
\left(\vec{\sigma}^{(i)}+\vec{\sigma}^{(j)}\right)\right]\cdot
\frac{\vec{k}}{\vec{k}^{\,2}+m_{\pi}^{2}}
\nonumber \\
& & -\frac{i}{2}\left[\bar{C}_{1}
+\bar{C}_{2}\,\boldtau^{(i)}\cdot\boldtau^{(j)}\right]
\left(\vec{\sigma}^{\,(i)}-\vec{\sigma}^{\,(j)}\right)\cdot\vec{k}\ ,
\label{eq:HPVTV_LO}
\end{eqnarray}
where $\vec{\sigma}^{(n)}/2$ ($\boldtau^{(n)}/2$) is the spin
(isospin) vector of nucleon $n$, and $\vec{k}=\vec{p}_{i}-\vec{p}_{i}^{\; '}$
is the momentum transferred from nucleon $i$. In this expression, at
LO $\bar{g}_{0}$ originates from $\tb$-term, qCEDM, and $\chi$I sources;
$\bar{g}_{1}$ from qCEDM and $\chi$I sources; and $\bar{C}_{i}$ from
$\chi$I sources only. The pion-exchange parts are well known (for example,
Refs. \cite{HaxHenley83,Khrip,Liu04}), while the contact interactions
incorporate all other $\slashPT$ effects of short-range, such as
single exchanges of the mesons $\omega$ and $\eta$ ($\bar{C}_{1}$) and
$\rho$ ($\bar{C}_{2}$) \cite{TVpotential}.

\subsection{Currents}
\label{genericcurrents}

As we argued above, only one-body currents are necessary at LO. For
the $\tb$ term, qCEDM, and $\chi$I sources we need the $PT$ current coming
from the proton charge in Eq. \eqref{LagStrong012}, 
\begin{eqnarray}
J_{PT}^{0} & = & \frac{e}{2}\left(1+\tau_{3}^{(i)}\right)\ ,
\label{charge}\end{eqnarray}
where $\boldtau^{(i)}/2$ is the isospin of the nucleon that couples
to the one-body current.

For the qEDM and $\chi$I sources we need as well
the $\slashPT$ current originating
from the nucleon EDMs, 
\begin{eqnarray}
J_{\slashPTsub}^{0} & = & - i\left(\bar{d}_{0}+\bar{d}_{1}\tau_{3}^{(i)}\right)\,
\vec{\sigma}^{(i)}\cdot\vec{q}\ ,
\label{nucEDM}
\end{eqnarray}
 where $\vec{\sigma}^{(i)}/2$ is the spin of the nucleon that interacts
with the photon and $\vec{q}\,$ is the outgoing photon momentum.

\section{Ingredients: nuclei with $N=Z$}
\label{isoscalar}

Although the power counting discussed above holds for general light
nuclei, it is possible that a diagram, which is expected to be LO,
does not contribute to the EDMs of certain systems. For nuclei of
equal neutron and proton number, $N=Z$, \textit{i.e.} the third component
of isospin $I_{3}=0$, an insertion of the isoscalar $\slashPT$ potential in
combination with the LO one-body $PT$ current, \textit{i.e.} Eq.
(\ref{charge}), does not contribute to the EDM \cite{HaxHenley83}.
To see this, consider the EDM operator resulting from the LO one-body
$PT$ current, which takes the simple expression 
\begin{equation}
\vec{D}_{PT}^{(1)}=
  \frac{e}{2}\,\sum_{i=1}^{A}\,\left(1+\tau_{3}^{(i)}\right)\,\vec{\xi}_{i}
= \frac{e}{2}\,\sum_{i=1}^{A}\,\tau_{3}^{(i)}\,\vec{\xi}_{i}
\label{eq:C1_TC_1B}
\end{equation}
in intrinsic coordinates $\vec{\xi}_{i}$ with $\sum_{i=1}^{A}\vec{\xi}_{i}=0$.
Since this operator is isovector, \textit{i.e.} $\Delta I=1$, and
conserves $I_{3}$, \textit{i.e.} 
$\Delta I_{3}=0$, it can only yield a non-vanishing moment when the
nuclear state of a 
$(I\,,\, I_{3}=0)$ nucleus acquires some parity admixture with isospin
$(I^{'}=I\pm1\,,\, I_{3}^{'}=0)$. Therefore, one needs isovector
components in $V_{\slashPTsub}$ to induce such admixture. The above argument holds in the non-relativistic limit.

This observation is of no concern for sources where there are other
contributions at the same order as those contributions that vanish.
The nuclear EDM is then simply dominated by the non-vanishing LO terms.
For the $\tb$ term, however, the LO contribution consists only of
an insertion of the isoscalar $\slashPT$ potential, such that, for
$N=Z$ nuclei, we need to go further down in power counting 
to find the dominant EDM contributions.

\subsection{Power counting}

Because the formally leading diagram (c) of Fig. \ref{classes} vanishes
for $N=Z$ in the $\tb$-term case when both the $PT$ one-body current
and the $\slashPT$ two-body potential are used, let us first consider
corrections in this diagram. It turns out that NLO corrections to
both the $\slashPT$ potential \cite{TVpotential} and $PT$ one-body
current vanish, and the first corrections we need to account for are
at NNLO. By looking at the scaling of the LECs for the $\tb$ term
in Eq. (\ref{NDAtheta}) and the power counting for the classes of
diagrams in Fig. \ref{classes}, 
we then conclude that the first non-vanishing contributions can come
from all classes of diagrams: the LO nucleon EDMs in diagram (a),
the LO $\slashPT$ two-body currents in diagram (b), the NNLO $\slashPT$
two-body potential \textit{or} the NNLO $PT$ one-body current in
diagram (c), and the LO $PT$ two-body currents with the LO $\slashPT$
two-body potential in diagram (d).

For the other sources only parts of the LO contributions given in
the previous section remain. For qCEDM and $\chi$I 
sources we need the $\slashPT$ potential from $\bar{g}_{1}$ OPE.
For qEDM and 
$\chi$I sources we also need the isoscalar short-range contribution to
the nucleon EDM.

\subsection{$P$- and $T$-odd potential}
\label{subLOpot}

For qCEDM and $\chi$I sources 
we can use the same potential as in the generic case, but the $\bar{g}_{0}$
and $\bar{C}_{1,2}$ terms will not contribute. We do not require
a $\slashPT$-potential for qEDM. For the $\tb$ term we need the
NNLO $\slashPT$ potential calculated in Ref. \cite{TVpotential}.
At this order further isoscalar terms appear, which also will not
contribute. Thus we need here only the following terms:
\begin{eqnarray}
V_{\slashPTsub}(\vec{k},\vec{K}, \vec{P}) & = & 
\frac{i}{2F_{\pi}^{2}}\left[ \left(g_{A}\bar{g}_{1}
-\frac{\bar{g}_{0}\beta_{1}}{2}\right)
\left(\tau_{3}^{(i)}+\tau_{3}^{(j)}\right)\,
\left(\vec{\sigma}^{(i)}-\vec{\sigma}^{(j)}\right)
\right.
\nonumber \\
& & \left.
+\left(g_{A}\bar{g}_{1}+\frac{\bar{g}_{0}\beta_{1}}{2}\right)
\left(\tau_{3}^{(i)}-\tau_{3}^{(j)}\right)\,
\left(\vec{\sigma}^{(i)}+\vec{\sigma}^{(j)}\right)\right]
\cdot\frac{\vec{k}}{\vec{k}^{\,2}+m_{\pi}^{2}}
\nonumber \\
& & +i\,\frac{\bar{g}_{0}g_{A}}{3F_{\pi}^{2}}
\left[\breve{\delta}m_{\pi}^{2}-\delta^{}m_{\pi}^{2}
-\frac{(\breve{\delta}m_{\pi}^{2})^{2}}{\vec{k}^{2}+m_{\pi}^{2}}
-\delta m_{N}^{2}\right]
\left(3\,\tau_{3}^{(i)}\tau_{3}^{(j)}-\boldtau^{(i)}\cdot\boldtau^{(j)}\right)
\nonumber \\
& & \times\left(\vec{\sigma}^{(i)}-\vec{\sigma}^{(j)}\right)\cdot\frac{\vec{k}}
 {(\vec{k}^{\,2}+m_{\pi}^{2})^{2}}
+\frac{\bar{g}_{0}g_{A}}{F_{\pi}^{2}}\frac{\delta m_{N}}{m_{N}}
\left(\boldtau^{\,(i)}\times\boldtau^{\,(j)}\right)_{3} 
\nonumber\\
& & \times\left[\left(\vec{\sigma}^{(i)}+\vec{\sigma}^{(j)}\right)\cdot\vec{K}
+\left(\vec{\sigma}^{(i)}-\vec{\sigma}^{(j)}\right)\cdot
\left(\frac{\vec{P}}{2}+\frac{(\vec{P}\cdot\vec{k})\,\vec{k}}
{\vec{k}^{\,2}+m_{\pi}^{2}}\right)\right]\frac{1}{\vec{k}^{\,2}+m_{\pi}^{2}}\ ,
\label{eq:HPVTV_N2LO}
\end{eqnarray}
where $\vec{P}=\vec{p}_{i}+\vec{p}_{j}$ is the center-of-mass (CM)
momentum of the nucleon pair and 
$\vec{K}=(\vec{p}_{i}+\vec{p}_{i}^{\; '}-\vec{p}_{j}-\vec{p}_{j}^{\;'})/4$.
The first two terms originate in one-pion exchange with $\bar{g}_{1}$
instead of $\bar{g}_{0}$ or with $\beta_{1}$ instead of $g_{A}$.
The next term arises from isospin breaking in the pion and nucleon
masses, and it is very small \cite{TVpotential}. The last term is due
to isospin breaking in the pion-nucleon vertex.
The potential also includes $1/m_N^2$ corrections \cite{TVpotential},
which we do not include here for the reasons given below.

\subsection{Currents}
\label{subLOcurrents}

For the same reasons that require the NNLO $\slashPT$ potential we
also need the NNLO $PT$ one-body electric current, to be used with
the $\bar{\theta}$-term LO potential. Again we do not bother with
terms that give a vanishing contribution for $N=Z$ nuclei. The only
remaining correction from Eq. \eqref{LagStrong012} is given by 
\begin{eqnarray}
J_{PT}^{0} & = & 
-\frac{ie}{16m_{N}^{2}}\varepsilon^{lmn}\sigma^{(i)l}q^{m}
\left(p_{i}+p_{i}'\right)^{n}
\left[1+2\kappa_{0}+\left(1+2\kappa_{1}\right)\tau_{3}^{(i)}\right]\ ,
\end{eqnarray}
which agrees with Ref. \cite{Manohar}. 
Here $\vec{p}_i$ ($\vec{p}_i^{\; '}$) is the momentum of the nucleon 
that couples to the photon before (after) interaction.

We also need two-body currents, both $PT$ and $\slashPT$. We use
incoming momenta $\vec{p}_{i}=\vec{P}/2+\vec{p}$ and 
$\vec{p}_{j}=\vec{P}/2-\vec{p}$
and outgoing momenta $\vec{p}_{i}^{\;'}=\vec{P}^{\; '}/2+\vec{p}\,'$ and 
$\vec{p}_{j}^{\;'}=\vec{P}^{\;'}/2-\vec{p}\,'$.
The photon momentum $\vec{q}=\vec{P}-\vec{P}^{\;'}$ is outgoing. 
For convenience
we introduce $\vec{k}=\vec{p}-\vec{p}\,'$ as before, 
$\vec{K}=(\vec{p}+\vec{p}\,')/2$,
and $\vec{P}_{t}=(\vec{P}+\vec{P}^{\;'})/2$. In the evaluation of the
currents at the order we are interested we can use the nucleon on-shell
relation $p_{n}^{0}=\vec{p}_{n}^{\,2}/2m_{N}$, or alternatively 
$k^{0}=(\vec{P}_{t}\cdot\vec{k}-\vec{q}\cdot\vec{K})/2m_{N}$.

\begin{figure}[t]
\centering \includegraphics[scale = 0.6]{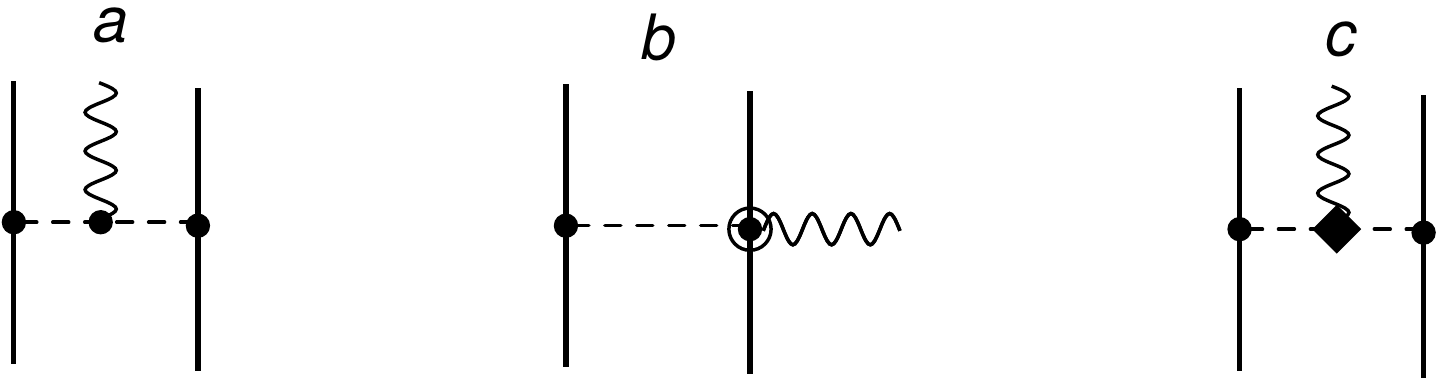}
\caption{Diagrams contributing to the $PT$ two-nucleon electric current. Solid,
dashed, and wavy lines represent nucleons, pions, and photons. A diamond
marks an isospin-breaking $PT$ interaction and the other vertices
isospin-conserving $PT$ interactions: leading (filled circles) and
subleading (circled circles). Only one topology per diagram is shown.}
\label{currentsTeven}
\end{figure}

The relevant diagrams for the LO two-body $PT$ electric current,
used again in combination with the LO $\slashPT$ two-body potential,
are shown in Fig. \ref{currentsTeven}. All interactions come from
the $PT$ Lagrangian, Eq. \eqref{LagStrong012}. In momentum space
the current reads 
\begin{eqnarray}
J_{PT,a}^{0} & = & + \frac{2ieg_{A}^{2}}{\Fp^{2}}
\left(\boldtau^{(i)}\times\boldtau^{(j)}\right)_{3}\, k^{0}
\frac{[\vec{\sigma}^{(i)}\cdot(\vec{k}+\vec{q}/2)]
[\vec{\sigma}^{(j)}\cdot(\vec{k}-\vec{q}/2)]}
{[(\vec{k}+\vec{q}/2)^{2}+\mpi^{2}][(\vec{k}-\vec{q}/2)^{2}+\mpi^{2}]}\ ,
\nonumber \\
J_{PT,b}^{0} & = & - \frac{ieg_{A}^{2}}{2\Fp^{2}m_{N}}
\left(\boldtau^{(i)}\times\boldtau^{(j)}\right)_{3}
\nonumber \\
& & \times\left[\frac{[\vec{\sigma}^{(i)}\cdot(\vec{P}_{t}+2\vec{K})]
[\vec{\sigma}^{(j)}\cdot(\vec{k}-\vec{q}/2)]}
{(\vec{k}-\vec{q}/2)^{2}+\mpi^{2}}
+\frac{[\vec{\sigma}^{(j)}\cdot(\vec{P}_{t}-2\vec{K})]
[\vec{\sigma}^{(i)}\cdot(\vec{k}+\vec{q}/2)]}{(\vec{k}+\vec{q}/2)^{2}+\mpi^{2}}
\right]\ ,
\nonumber \\
J_{PT,c}^{0} & = & -\frac{2eg_{A}^{2}}{\Fp^{2}}\delta m_{N}
\left(\boldtau^{(i)}\cdot\boldtau^{(j)}-\tau_{3}^{(i)}\tau_{3}^{(j)}\right)
\frac{[\vec{\sigma}^{(i)}\cdot(\vec{k}+\vec{q}/2)]
[\vec{\sigma}^{(j)}\cdot(\vec{k}-\vec{q}/2)]}
{[(\vec{k}+\vec{q}/2)^{2}+\mpi^{2}][(\vec{k}-\vec{q}/2)^{2}+\mpi^{2}]} \ .
\label{eq:J0_PCTC_2B}
\end{eqnarray}

%%%%%%%%%%%%%%%%%%%%%
%
%
%%%%%%%%%%%%%%%%%%%%%

\begin{figure}[t]
\centering \includegraphics[scale = 0.6]{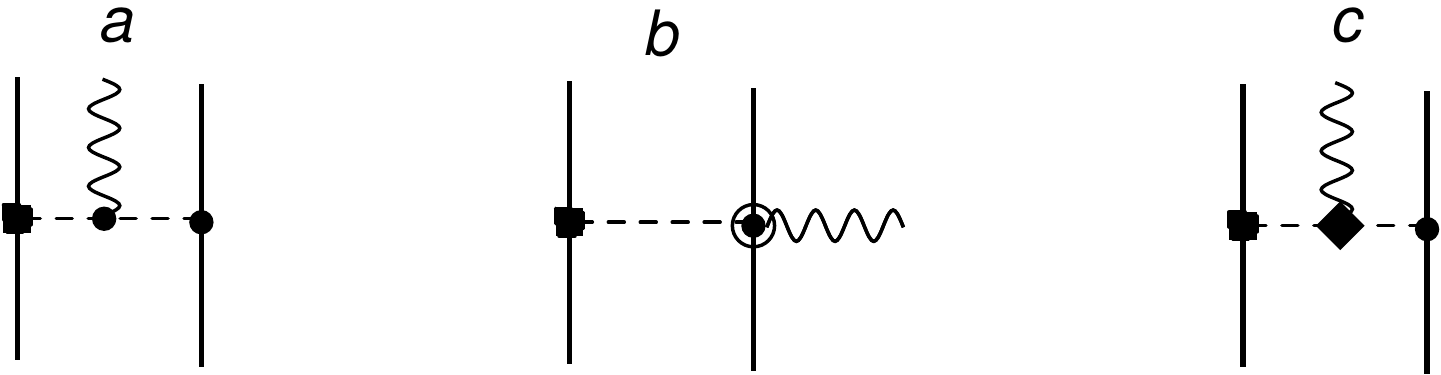}
\caption{Diagrams contributing to the $\slashPT$ two-nucleon electric current.
A square marks a $\slashPT$ interaction;
other notation as in Fig. \ref{currentsTeven}.
Only one topology per diagram is shown.}
\label{currentsTodd} 
\end{figure}

We also need to include the LO two-nucleon $\slashPT$ electric current.
The diagrams contributing to this current are shown in Fig. \ref{currentsTodd}.
Here $PT$ interactions come from the $PT$ Lagrangian, Eq. \eqref{LagStrong012},
and the $\slashPT$ interaction is the $\bar{g}_{0}$ vertex in the
$\slashPT$ Lagrangian, Eq. \eqref{chiLag123}. The current is given by 
\begin{eqnarray}
J_{\slashPTsub,a}^{0} & = & + \frac{2eg_{A}\bar{g}_{0}}{\Fp^{2}}
\left(\boldtau^{(i)}\times\boldtau^{(j)}\right)_{3}\, k^{0}
\frac{(\vec{\sigma}^{(i)}+\vec{\sigma}^{(j)})\cdot\vec{q}/2
+(\vec{\sigma}^{(i)}-\vec{\sigma}^{(j)})\cdot\vec{k}}
{[(\vec{k}+\vec{q}/2)^{2}+\mpi^{2}][(\vec{k}-\vec{q}/2)^{2}+\mpi^{2}]}\ ,
\nonumber \\
J_{\slashPTsub,b}^{0} & = & -\frac{eg_{A}\bar{g}_{0}}{2\Fp^{2}m_{N}}
\left(\boldtau^{(i)}\times\boldtau^{(j)}\right)_{3}
\left[\frac{\vec{\sigma}^{(i)}\cdot(\vec{P}_{t}+2\vec{K})}
{(\vec{k}-\vec{q}/2)^{2}+\mpi^{2}}-\frac{\vec{\sigma}^{(j)}
\cdot(\vec{P}_{t}-2\vec{K})}{(\vec{k}+\vec{q}/2)^{2}+\mpi^{2}}\right]\ , \\
J_{\slashPTsub,c}^{0} & = & +\frac{2ieg_{A}\bar{g}_{0}}{\Fp^{2}}\delta m_{N}
\left(\boldtau^{(i)}\cdot\boldtau^{(j)}-\tau_{3}^{(i)}\tau_{3}^{(j)}\right)
\frac{(\vec{\sigma}^{(i)}+\vec{\sigma}^{(j)})\cdot\vec{q}/2
+(\vec{\sigma}^{(i)}-\vec{\sigma}^{(j)})\cdot\vec{k}}
{[(\vec{k}+\vec{q}/2)^{2}+\mpi^{2}][(\vec{k}-\vec{q}/2)^{2}+\mpi^{2}]} \ . 
\nonumber
\label{eq:J0_PVTV_2B}
\end{eqnarray}

%%%%%%%%%%%%%%%%%%%%%
%
%
%%%%%%%%%%%%%%%%%%%%%

\section{EDM of the Deuteron}
\label{Deuteron}

We are now in position to calculate the EDM of the deuteron, which
provides the simplest example of an $N=Z$ nucleus. The ground state
of the deuteron is mainly a $^{3}S_{1}$ state. The deuteron obtains
a $^{1}P_{1}$ component after a $\bar{g}_{0}$ pion exchange or an
insertion of $\bar{C}_{1,2}$. Since the LO $PT$ one-nucleon current
is spin independent, it cannot bring the deuteron wavefunction from
$^{1}P_{1}$ to $^{3}S_{1}$, and therefore these contributions vanish
for the deuteron, as anticipated on more general grounds in the previous
section.

The deuteron EDM has been studied before in the meson-exchange picture
\cite{SFK84,Avi85,Khrip,Liu04,Afn10}, with various degrees of sophistication
in the treatments of the $P$- and $T$-conserving interaction $H_{PT}$.
Using modern high-quality phenomenological potentials \cite{Sto94,Wir95},
Ref.~\cite{Liu04} found that the model dependence of $H_{PT}$ is
rather small for a deuteron EDM generated by the OPE
sector of the $\slashPT$ interaction. The detailed study in Ref.~\cite{Afn10}
confirmed this point. Since our new EFT scheme shows that the
leading-order contribution from various $\slashPT$ sources to the
deuteron EDM also comes from the long-range terms in $V_{\slashPTsub}$,
we take advantage of the existing calculation scheme of Ref.~\cite{Liu04}
to obtain wave functions $|\Psi_{^{2}\mathrm{H}}\rangle$ and 
$|\widetilde{\Psi}_{^{2}\mathrm{H}}\rangle$.
The calculation is performed in coordinate space
using the $\slashPT$ potentials and currents
from Appendices \ref{Vcoord} and \ref{Fourier}, respectively.
Of course, a fully consistent treatment would involve using the 
$PT$ interaction $H_{PT}$ derived from the complete chiral Lagrangian, instead
of a phenomenological potential. At present, unfortunately, such a consistent
potential does not exist beyond LO \cite{morethanweinberg}. It would
include relativistic corrections as well, which are absent in the
phenomenological potentials we use. For this reason, we neglect relativistic
corrections in the $\slashPT$ potential and currents as well. We
expect that the results from a fully consistent calculation will not
deviate significantly from the results we obtain here.
The numbers below correspond to
the Argonne $v_{18}$ potential \cite{Wir95}, but results for the 
Reid93 and Nijmegen II potentials \cite{Sto94} agree within $5$\%.
This is 
less than the error of order $m_\pi/\MQCD\sim 20$\% 
intrinsic to $\chi$PT in lowest order.

The simplest contribution to the deuteron EDM comes
from the constituent EDMs.
The LO $J_{\slashPTsub}^{0}$, given in Eq. (\ref{nucEDM}), yields a
one-body EDM operator 
\begin{equation}
\vec{D}_{\slashPTsub}^{(1)}= \sum_{i=1}^{A}\,
\left(\bar{d}_{0}+\bar{d}_{1}\,\tau_{3}^{(i)}\right) \,
\vec{\sigma}^{(i)} \ .
\label{eq:C1_TV_1B}
\end{equation}
For the deuteron, an isoscalar ($I=0$) and spin-triplet ($S=1$)
state, one simply gets 
\begin{equation}
\frac{1}{\sqrt{6}}\,
\left\langle \Psi_{^{2}\mathrm{H}}\left|\left|\vec{D}_{\slashPTsub}^{(1)}
\right|\right|\Psi_{^{2}\mathrm{H}}\right\rangle 
=d_n+d_p \ .
\end{equation}

In order for $\vec{D}_{PT}^{(1)}$, a purely isovector operator as
discussed earlier,
to yield a non-zero contribution in the deuteron,
it is obvious that the parity admixture 
$|\widetilde{\Psi}_{^{2}\mathrm{H}}\rangle$
has to be a $^{3}P_{1}$ state.
Among the various terms in the LO $\slashPT$ potential, 
Eq. (\ref{eq:HPVTV_LO}), only
the one with the isospin-spin operator 
$(\tau_{3}^{(1)}-\tau_{3}^{(2)})(\vec{\sigma}^{(1)}+\vec{\sigma}^{(2)})$ 
can contribute. The result is 
\begin{equation}
\frac{2}{\sqrt{6}}\,\left\langle \Psi_{^{2}\mathrm{H}}\left|\left|
\vec{D}_{PT}^{(1)}\right|\right|\widetilde{\Psi}_{^{2}\mathrm{H}}(^3P_1)
\right\rangle 
=-0.19\,\frac{\bar{g}_{1}}{\Fp}\; e\,\mathrm{fm}\ .
\label{eq40}
\end{equation}
However, when it comes to the $\tb$ term, because
the LO contribution vanishes as argued in the previous section, the leading
contribution is in fact NNLO. Among the higher-order interactions
identified in Section \ref{subLOpot}, the terms with coupling constants
$\left(g_{A}\bar{g}_{1}+\bar{g}_{0}\beta_{1}/2\right)$ and 
$g_{A}\bar g_0 \delta m_N$ in Eq. (\ref{eq:HPVTV_N2LO})
can contribute, by isospin and spin selection rules. 
Except for the coupling constants, the operator structures of the former are
the same as the one in Eq. (\ref{eq:HPVTV_LO}), 
so the matrix element can simply be obtained by replacing
\begin{equation}
\bar{g}_{1}\rightarrow\bar{g}_{1}+\frac{\beta_{1}}{2g_{A}}\bar{g}_{0} 
\end{equation}
in Eq. \eqref{eq40}.
Combining this with the contribution from the isospin-breaking pion-nucleon
vertex,
we find the matrix element 
\begin{equation}
\frac{2}{\sqrt{6}}\,\left\langle \Psi_{^{2}\mathrm{H}}\left|\left|
\vec{D}_{PT}^{(1)}\right|\right|\widetilde{\Psi}_{^{2}\mathrm{H}}(^3P_1)
\right\rangle 
= -\left[0.19\,\left( \frac{\bar{g}_{1}}{\Fp}
+\frac{\beta_{1}}{2g_{A}}\frac{\bar{g}_{0}}{\Fp}\right)
+ 5.8 \cdot 10^{-4}\, \frac{\bar{g}_{0}}{\Fp}\right] e\,\mathrm{fm}
\end{equation}
for the $\tb$ term.

For the $\tb$ term, there are in addition NNLO currents to be taken
into account. For the $\slashPT$ currents, as the corresponding EDM
operators are sandwiched between two isoscalar states, they must be
isoscalar to contribute. Among the NNLO $\slashPT$ currents identified
in Section \ref{subLOcurrents}, 
only the third current in Eq. (\ref{eq:J0_PVTV_2B}),
$J_{\slashPTsub,c}^{0}$, meets the requirement and leads to a
two-body EDM operator
(see Appendix \ref{Fourier})
\begin{equation}
\vec{D}_{\slashPTsub}^{(2)}=-e\,\frac{g_{A}\,\bar{g}_{0}}{F_{\pi}^{2}}\,
\delta m_{N}\,
\left(\boldtau^{(1)}\cdot\boldtau^{(2)}-\tau^{(1)}_{3}\,\tau^{(2)}_{3}\right)
\left[
\vec{\sigma}^{(1)}\cdot\vec{\nabla}_{1}
+\vec{\sigma}^{(2)}\cdot\vec{\nabla}_{2}\,,
\,(\vec{x}_{1}+\vec{x}_{2})\,
\frac{e^{-m_{\pi}|\vec{x}_{1}-\vec{x}_{2}|}}{8\,\pi\, m_{\pi}}\right]
\label{D2slashPT}
\end{equation}
in terms of the positions $\vec{x}_{1}$ and $\vec{x}_{2}$
of the two nucleons and the derivatives 
$\vec{\nabla}_{1}$ and $\vec{\nabla}_{2}$ with respect to them.
This results in the matrix element
\begin{equation}
\frac{1}{\sqrt{6}}\,\left\langle \Psi_{^{2}\mathrm{H}}\left|\left|
\vec{D}_{\slashPTsub}^{(2)}\right|\right|
\Psi_{^{2}\mathrm{H}}\right\rangle =1.1\cdot10^{-3}\,\frac{\bar{g}_{0}}{\Fp}\;
e\,\mathrm{fm}\label{resultToddcurrent}
\end{equation}
for the deuteron EDM.
The contributions of two-body $PT$ currents to the EDM have 
again to be coupled with
the parity admixture generated by the LO $V_{\slashPTsub}$, which is
purely isoscalar when $\tb$ is the $\slashPT$ source. 
The only $PT$ current with an isoscalar component, among those
identified in Section \ref{subLOcurrents}, 
is the third current in Eq. (\ref{eq:J0_PCTC_2B}), $J_{PT,c}^{0}$.
It gives a two-body EDM operator 
\begin{equation}
\vec{D}_{PT}^{(2)}=-e\,\frac{g_{A}^{2}}{F_{\pi}^{2}}\,\delta m_{N}\,
\left(\boldtau^{(1)}\cdot\boldtau^{(2)}-\tau^{(1)}_{3}\,\tau^{(2)}_{3}\right)
\left[(\vec{\sigma}^{(1)}\cdot\vec{\nabla}_{1})
(\vec{\sigma}^{(2)}\cdot\vec{\nabla}_{2})\,,
\,(\vec{x}_{1}+\vec{x}_{2})\,\frac{e^{-m_{\pi}|\vec{x}_{1}-\vec{x}_{2}|}}
{8\,\pi\, m_{\pi}}\right]\ .
\label{D2PT}
\end{equation} 
Since the isoscalar parity admixture 
$|\widetilde{\Psi}_{^{2}\mathrm{H}}\rangle$
can only be a $^{1}P_{1}$ state, this current gives a matrix element
\begin{equation}
\frac{2}{\sqrt{6}}\,\left\langle \Psi_{^{2}\mathrm{H}}\left|\left|
\vec{D}_{PT}^{(2)}\right|\right|
\widetilde{\Psi}_{^{2}\mathrm{H}}(^{1}P_{1})\right\rangle =
-3.3\cdot10^{-4}\,\frac{\bar{g}_{0}}{\Fp}\; e\,\mathrm{fm} \ .
\label{resultTevencurrent}
\end{equation}

In total the deuteron EDM can be written as a function of three 
$\slashPT$ LECs,
\begin{equation}
d_{^{2}\mathrm{H}} = d_{p}+d_{n}
+\left[-0.19\, \frac{\bar{g}_{1}}{\Fp}
+\left(0.2 - 0.7\cdot 10^{2}\,\beta_{1}\right)
\cdot 10^{-3} \frac{\bar{g}_{0}}{\Fp}\; 
\right] e\,\mathrm{fm}\ ,
\label{dEDM}
\end{equation}
where $d_{p,n}$ should be included for $\tb$, qEDM, and $\chi$I;
$\bar{g}_{1}$ for $\tb$, qCEDM, and $\chi$I;
and $\bar{g}_{0}$ for $\tb$ only.

This result can be compared, for each of the sources, with the calculation
where OPE is treated perturbatively \cite{Vri11b}. 
For both qCEDM and qEDM the nonperturbative pion approach adopted
here agrees very well with the perturbative calculation.
In the case of the qCEDM,
it was also found that the deuteron EDM 
is dominated by $\bar{g}_{1}$ pion exchange 
and given by \cite{Vri11b} 
\begin{equation}
d_{^{2}\mathrm{H}}(\mathrm{qCEDM})|_{pert}=
-\frac{eg_{A}\bar{g}_{1}m_{N}}{6\pi\Fp^{2}\mpi}
\frac{1+\gamma/\mpi}{(1+2\gamma/\mpi)^{2}}
=-0.23\,\frac{\bar{g}_{1}}{\Fp}\; e\,\mathrm{fm}\ ,
\label{onebodyEDFF}
\end{equation}
where  $\gamma\simeq 45\;\mathrm{MeV}$ is the binding
momentum of the deuteron. This result agrees exactly with a zero-range
model \cite{Khrip} and is $22\%$ larger than the result from
the qCEDM calculation with nonperturbative OPE \cite{Liu04}
reproduced above,
\begin{equation}
d_{^{2}\mathrm{H}}(\mathrm{qCEDM}) = 
-0.19\, \frac{\bar{g}_{1}}{\Fp}\, e\,\mathrm{fm} \ .
\label{dEDMqCEDM}
\end{equation}
Since the estimated error in the perturbative calculation is of order 
$Q/M_{N\! N}\sim30\%$,
the calculations agree within their uncertainty. 
By power counting the contribution from $\bar d_0$ is expected to be suppressed by $\mpi^2/M_{\mathrm{QCD}}^2$ compared to Eq. (\ref{dEDMqCEDM}). From Eq. (\ref{NDAqCEDM}) we infer $ \bar d_0 \Fp/\bar g_1 = \Or( e \Fp/\MQCD^2)  \sim 0.03 \; e\, \mathrm{fm}$, implying that, in the case of qCEDM, the nucleon EDMs contribute at the $30$\% level to the deuteron EDM. This suppression is less than formally expected. If we assume the isoscalar nucleon EDM is saturated by its long-range part, Eq. (\ref{d0est}), the contribution is at the $10$\% level. In any case, the correction by the isoscalar nucleon EDM is of the order of the intrinsic $\chi$PT uncertainty $\mpi/\MQCD$, such that for the qCEDM the deuteron EDM at LO is given by Eq. (\ref{dEDMqCEDM}).

Likewise, for qEDM the
conclusions of Ref. \cite{Vri11b} do not change once 
we treat OPE nonperturbatively.
The deuteron EDM is in this case simply the sum of the neutron and
proton EDM, 
\begin{equation}
d_{^{2}\mathrm{H}}(\mathrm{qEDM})=2\bar{d}_{0}\ .
\label{dEDMqEDM}
\end{equation}

The comparison is more subtle for $\tb$ and $\chi$I $\slashPT$ sources.
For both of these sources, 
the deuteron EDM is expected in the perturbative-pion approach 
to be dominated by the isoscalar nucleon EDM, 
since pion exchange is further suppressed
in the $Q/M_{N\! N}$ expansion.
In the nonperturbative power counting
$\slashPT$ pion exchange is a dominant effect as well. 
In order to
compare the two effects
---nucleon EDMs and pion exchange--- 
in the nonperturbative calculation we can
look at the estimated scaling of the LECs.
For $\chi$I sources,
\begin{equation}
d_{^{2}\mathrm{H}}(\chi\mathrm{I}) = 2\bar{d}_{0}
-0.19\, \frac{\bar{g}_{1}}{\Fp}\, e\,\mathrm{fm} \ .
\label{dEDMchiI}
\end{equation}
{}From Eq. (\ref{NDATVCI}) 
we infer that
$F_\pi \bar{d}_{0}/\bar{g}_{1}=\Or(e F_\pi/\varepsilon\mpi^{2})
\sim 5 
\; e\, \mathrm{fm}$.
Thus, although formally $\bar{g}_{1}$ exchange is LO, 
because of a combination of $\varepsilon$ suppression 
and the relatively small factor of $0.19$ in Eq. \eqref{dEDM},
it actually is expected to
contribute only at the $\sim 5\%$ level to the deuteron EDM. 
For $\tb$ there are additional contributions from $\bar{g}_{0}$,
\begin{equation}
d_{^{2}\mathrm{H}}(\tb)  = 2\bar{d}_{0}
+\left[-0.19\, \frac{\bar{g}_{1}}{\Fp}
+\left(0.2-0.7\cdot 10^{2}\,\beta_{1}\right)\cdot 10^{-3}
\frac{\bar{g}_{0}}{\Fp}\right]  e\,\mathrm{fm}\ .
\label{dEDMtheta}
\end{equation}
The contributions from the $\slashPT$ and $PT$ two-body currents,
Eqs. \eqref{D2slashPT} and \eqref{D2PT} respectively, are 
of similar size.
The $N\!N$ data constraint 
\cite{isoviolphen} on $\beta_1$ shows that the contribution
from the $\slashPT$ potential is no larger, and 
the full $\bar{g}_{0}$ term is 
$\simle 0.9 \cdot 10^{-3} (\bar{g}_{0}/\Fp) \, e \, \mathrm{fm}$.
{}From Eq. \eqref{NDAtheta} we expect that 
$\bar{g}_{1}/\bar{g}_{0}=\Or(\varepsilon m_\pi^2/\MQCD^{2}) 
\sim 10^{-2}$,
so the $\bar{g}_{1}/F_\pi$ contribution should be comparable
to these small $\bar{g}_{0}/F_\pi$ contributions.
In contrast, we expect a larger weight from
the pion cloud around each nucleon,
which for  
$\bar{d}_{0}$ enters at NLO and gives Eq. \eqref{d0est}.
Thus again, although pion-exchange contributions
in the potential and currents 
are formally LO,  $\varepsilon$ suppression 
and relatively small numerical factors in the deuteron make them
likely no more than $\sim 10\%$ of the nucleon
EDM contribution.

The fact that pion-exchange contributions are expected
to be smaller in the deuteron than assumed in $\chi$PT power counting
confirms that the power counting of Ref. \cite{Vri11b}, 
where pion exchange comes in at NLO, works better for 
a loosely bound nucleus.
The $\chi$PT power counting should become
more accurate as we consider heavier, denser nuclei,
the simplest of which we tackle next.

\section{EDM of the Helion and the Triton}
\label{helium} 

In this section we investigate the EDMs of $^{3}$He and
$^{3}$H. No particular cancellations are expected,
so the framework of Section \ref{generic} applies.

The EDM of $^{3}$He was studied in Ref. \cite{Ste08}, where 
two $\slashPT$ mechanisms were considered:
nucleon EDMs and a  $\slashPT$ two-nucleon potential
containing the
most general non-derivative, single $\pi$-, $\rho$-, and $\omega$-meson
exchanges.
The nuclear wavefunction was 
calculated 
with the no-core shell model (NCSM) \cite{NCSM},
where a $PT$ nuclear potential is solved
within a model space made from appropriately
symmetrized combinations \cite{lithuania}
of $N_{max}$ harmonic-oscillator wavefunctions of 
frequency $\Omega$. 
In Ref. \cite{Ste08} both Argonne $v18$ \cite{Wir95}
and EFT-inspired \cite{eftinspiredNN} potentials, 
including the Coulomb interaction,
were used.
At large enough $N_{max}$ results become independent of $\Omega$.

Here we adapt this calculation to the $\slashPT$
ingredients from chiral EFT, and calculate the EDM of $^{3}$H
for the first time.
As argued in Section \ref{generic}, power counting for generic
light nuclei tells us that for all $\slashPT$ sources
of dimension up to six, the EDM is indeed expected to
come mostly from the nucleon EDM and from the two-nucleon 
$\slashPT$ potential, as assumed in Ref. \cite{Ste08}.
The only difference is that the EFT potential \eqref{eq:HPVTV_LO}
contains, in addition to OPE, also two LECs ($\bar{C}_{1}$ and $\bar{C}_{2}$) 
representing shorter-range interactions. 
This potential in coordinate space is given in Appendix \ref{Vcoord}.
The OPE terms were included in Ref. \cite{Ste08}, while
$\bar{C}_{1}$ and $\bar{C}_{2}$ can be thought of as originating 
from, respectively, $\omega$ and $\rho$ exchanges, also considered there.
The relation can be made quite explicit
if we choose to regularize the delta functions with Yukawa functions,
following a strategy successfully employed before
to study the effects of the EFT $\slashP T$ potential \cite{CPonP}:
\begin{eqnarray}
\frac{m_1^2 \bar{C}_{1}}{4\pi r}e^{-m_1 r}  &\to&
\bar{C}_{1}\delta^{(3)}(\vec{r}\,) \ ,
\\
\frac{m_2^2 \bar{C}_{2}}{4\pi r}e^{-m_2 r}  &\to& 
\bar{C}_{2}\delta^{(3)}(\vec{r}\,) \ ,
\end{eqnarray}
as $m_{1,2}\to \infty$.
When $m_1=m_\omega$ ($m_2=m_\rho$) and $\bar{C}_{1}$ ($\bar{C}_{2}$) 
is an appropriate combination of $\omega$ ($\rho$) couplings 
\cite{TVpotential},
the expressions on the left-hand side coincide
with those in Ref. \cite{Ste08}.
Here we recalculate these contributions for values of $m_{1,2}$
up to $2.5$ GeV.
For uniformity with Section \ref{Deuteron}
we again display numbers obtained with the Argonne $v18$
potential. In Ref. \cite{Ste08} it was found that for helion the contributions
from nucleon EDMs $(\bar d_{0,1}$) and from pion exchange ($\bar g_{0,1}$)
change with $PT$ potential by no more than $\sim 25$\%. 
We have verified that the same is true for triton.
Unfortunately the situation is different for the short-range two-body
contributions ($\bar C_{1,2}$), which are much more sensitive to
the $PT$ potential, as we discuss shortly.

The nucleon EDM contributions are found to be
\begin{eqnarray}
\frac{1}{\sqrt{6}}\,
\left\langle \Psi_{^{3}\mathrm{He}}\left|\left|\vec{D}_{\slashPTsub}^{(1)}
\right|\right|\Psi_{^{3}\mathrm{He}}\right\rangle 
&=&0.88\, d_n - 0.047\, d_p \ ,
\\
\frac{1}{\sqrt{6}}\,
\left\langle \Psi_{^{3}\mathrm{H}}\left|\left|\vec{D}_{\slashPTsub}^{(1)}
\right|\right|\Psi_{^{3}\mathrm{H}}\right\rangle 
&=&-0.050\, d_n + 0.90\, d_p \ .
\label{3NfromN}
\end{eqnarray}
As expected, the helion (triton) EDM is mostly sensitive to the
neutron (proton) EDM \cite{Ste08}.

%%%%%%%%%%%%%%%%%%%%%
%
\begin{figure}[t]
\includegraphics[scale = 1]{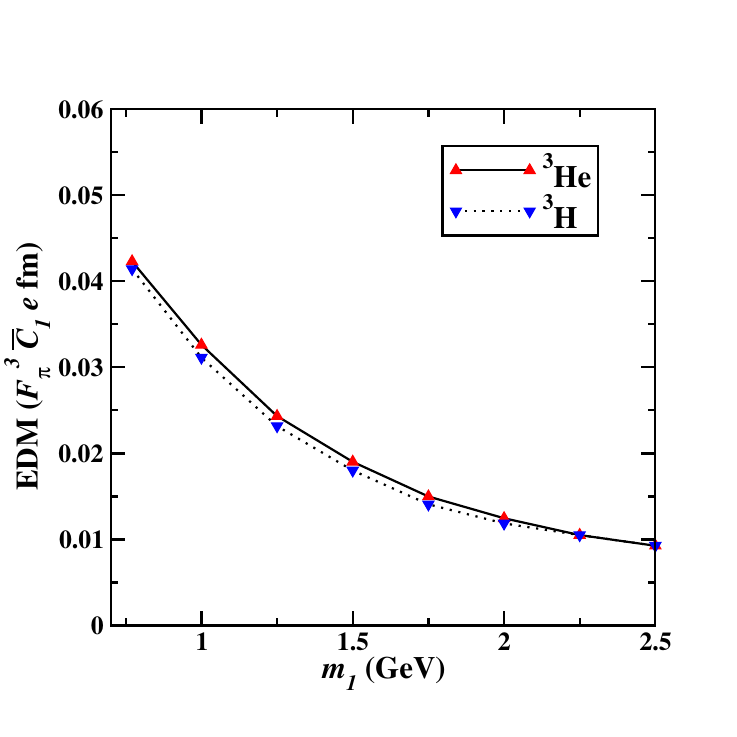}
\includegraphics[scale = 1]{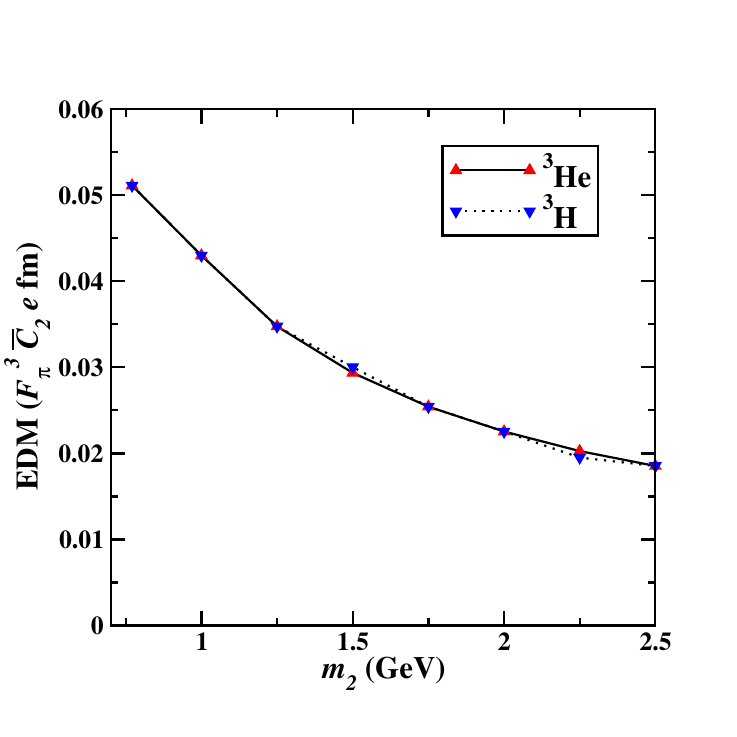}
\caption{Magnitude of the tri-nucleon EDMs in units
of $F_\pi^3 \bar{C}_i e$ fm, as function of the regulator mass in GeV:
$i=1$ (left panel) and $i=2$ (right panel).
The solid (dashed) curve is for helion (triton).}
\label{shortrangeplots}
\end{figure}
%
%%%%%%%%%%%%%%%%%%%%%

For the contribution from the $\slashPT$ potential, 
our results for triton are very similar in magnitude to 
those for helion, in the case of OPE already obtained in Ref. \cite{Ste08}.
The contribution of $\bar{C}_{1,2}$ as a function of $m_{1,2}$
is given in Fig. \ref{shortrangeplots} for Argonne $v18$.
For each regulator mass, we perform calculations
at four values of $\Omega=20, 30, 40, 50$ MeV, up to $N_{max}=50$.
We observe convergence and estimate a 10\% error from the spread
of results with $\Omega$.
(See Fig. 1 of Ref. \cite{Ste08} for a generic convergence pattern.)
As it can be seen from  Fig. \ref{shortrangeplots}, the results become 
approximately $m_{1,2}$ independent at large masses, implying that
$\bar{C}_{1,2}$ approach constants in this limit.
Results are very different for the EFT-inspired potential.
Within the region of masses studied, we found 
an approximately linear dependence on the regulator mass,
always larger in magnitude than for Argonne $v18$. 
While for $m_1=m_\omega$ and $m_2=m_\rho$ the contributions
to the tri-nucleon EDMs differ by a factor $\sim 2$ \cite{Ste08},
the difference first increases and then decreases as $m_{1,2}$ increases,
but it is still a factor of $\sim 5$ at $2.5$ GeV.
The linear regulator dependence could indicate
a different running of $\bar{C}_{1,2}$, or simply a very slow
convergence.
However, calculations with this potential are computationally more intensive
and we have been limited to $N_{max}=40$, which increases the error.
In any case, there is clearly a much stronger dependence of
these short-range contributions on the potential,
and more solid numbers have to await a fully consistent calculation.
We quote here the Argonne $v18$ numbers at 2.5 GeV, but we
emphasize that they represent only an order of magnitude estimate.
We obtain
\begin{eqnarray}
\frac{2}{\sqrt{6}}\,\left\langle \Psi_{^{3}\mathrm{He}}\left|\left|
\vec{D}_{PT}^{(1)}\right|\right|\widetilde{\Psi}_{^{3}\mathrm{He}}
\right\rangle 
&=&\left(-0.15\frac{\bar{g}_{0}}{\Fp}
-0.28\frac{\bar{g}_{1}}{\Fp}
-0.01 F_{\pi}^3 \bar{C}_{1}
+0.02 F_{\pi}^3 \bar{C}_{2}\right) e\, \textrm{fm} \ ,
\\
\frac{2}{\sqrt{6}}\,\left\langle \Psi_{^{3}\mathrm{H}}\left|\left|
\vec{D}_{PT}^{(1)}\right|\right|\widetilde{\Psi}_{^{3}\mathrm{H}}
\right\rangle 
&=&
\left(0.15\frac{\bar{g}_{0}}{\Fp}-0.28\frac{\bar{g}_{1}}{\Fp}
+ 0.01 F_{\pi}^3 \bar{C}_{1}
- 0.02 F_{\pi}^3 \bar{C}_{2}\right) e\, \textrm{fm} \ .
\label{3Nfrom2N}
\end{eqnarray}

In total, then, 
as anticipated in Sections \ref{ToddLag} and \ref{generic},
the EDMs of helion and triton (as
the EDMs of light nuclei in general) are functions of six $\slashPT$
LECs:
\begin{equation}
d_{^{3}\mathrm{He}} =  
0.88\, d_{n}-0.047\, d_{p}  
-\left(0.15\, \frac{\bar{g}_{0}}{\Fp}+0.28\, \frac{\bar{g}_{1}}{\Fp}
+0.01 \, F_{\pi}^3 \bar{C}_{1}
-0.02 \, F_{\pi}^3 \bar{C}_{2}\right) e\, \textrm{fm} 
\label{helionEDM}
\end{equation}
and
\begin{equation}
d_{^{3}\mathrm{H}} = 
-0.050\, d_{n}+0.90\, d_{p} 
+\left(0.15\,\frac{\bar{g}_{0}}{\Fp}-0.28\,\frac{\bar{g}_{1}}{\Fp}
+0.01 \, F_{\pi}^3 \bar{C}_{1} 
-0.02 \, F_{\pi}^3 \bar{C}_{2}\right) e\,\mathrm{fm} \ ,
\label{tritonEDM}
\end{equation}
where $\bar{g}_{0}$ applies for $\tb$, qCEDM, and $\chi$I;
$\bar{g}_{1}$ for qCEDM and $\chi$I; 
$d_{n,p}$ for qEDM and $\chi$I; 
and $\bar{C}_{1,2}$ for $\chi$I only. 

Only in the case of the qEDM do we expect the tri-nucleon EDMs to be dominated
by the nucleon EDMs. 
Not surprisingly,
the helion (triton) EDM should be approximately equal to the 
neutron (proton) EDM.
The nucleon EDM for the dimension-six sources
was calculated in Ref. \cite{Vri11a} and it was found that for qEDM the
EDMs were dominated by the short-range contributions
in Eqs. \eqref{dnLO} and \eqref{dpLO}.
In this case,
\begin{eqnarray}
d_{^{3}\mathrm{He}}(\mathrm{qEDM})  &=& 
0.83\,\bar{d}_{0}-0.93\,\bar{d}_{1} \ ,
\label{hEDMqEDM}
\\
d_{^{3}\mathrm{H}}(\mathrm{qEDM})  &=& 
0.85\,\bar{d}_{0}+0.95\,\bar{d}_{1} \ .
\label{tEDMqEDM}
\end{eqnarray}

For the $\tb$ term, on the other hand, the helion and triton EDMs depend 
at LO only on $\bar{g}_{0}$.
To check this statement we compare the LO contribution with the contribution
from the nucleon EDMs. 
If we assume the neutron and proton EDMs to be saturated by
their long-range part, that is, the chiral log in Eq. \eqref{d1},
Eq. \eqref{d1est} shows that the short-range term
is comparable to the pion-exchange contribution.
To be on the safe side, it seems better not to neglect the LO nucleon
EDMs for the $\tb$ term, even though the power counting tells us it should
be subleading; then
\begin{eqnarray}
d_{^{3}\mathrm{He}}(\tb)  &=& 
0.83\,\bar{d}_{0}-0.93\,\bar{d}_{1}
-0.15\, \frac{\bar{g}_{0}}{\Fp} \, e\,\mathrm{fm} \ ,
\label{hEDMtheta}\\
d_{^{3}\mathrm{H}}(\tb)  &=& 
0.85\,\bar{d}_{0}+0.95\,\bar{d}_{1} 
+0.15\, \frac{\bar{g}_{0}}{\Fp}\, e\,\mathrm{fm}\ .
\label{tEDMtheta}
\end{eqnarray}
This argument holds equally well for the qCEDM,
except that now also
$\bar{g}_{1}$ contributes:
\begin{eqnarray}
d_{^{3}\mathrm{He}}(\mathrm{qCEDM})  &=& 
0.83\,\bar{d}_{0}-0.93\,\bar{d}_{1} 
-\left(0.15\,\frac{\bar{g}_{0}}{\Fp}+0.28\,\frac{\bar{g}_{1}}{\Fp}\right)
 e\,\mathrm{fm} \ ,
\label{hEDMqCEDM}
\\
d_{^{3}\mathrm{H}}(\mathrm{qCEDM})  &=& 
0.85\,\bar{d}_{0}+0.95\,\bar{d}_{1} 
+\left(0.15\,\frac{\bar{g}_{0}}{\Fp}-0.28\,\frac{\bar{g}_{1}}{\Fp}\right)
 e\,\mathrm{fm} \ .
\label{tEDMqCEDM}
\end{eqnarray}

Finally in the case of $\chi$I, we expect the 
tri-nucleon EDM to consist
of $\bar{g}_{0,1}$ pion exchange, insertions of 
the $\slashPT$ short-range $N\!N$ interactions,
and the contributions from the nucleon EDMs. 
Similarly to the
qEDM, the nucleon EDM from $\chi$I is dominated by short-range contributions.
All six LECs contribute:
\begin{eqnarray}
d_{^{3}\mathrm{He}}(\mathrm{\chi I}) &=& 
0.83\,\bar{d}_{0}-0.93\,\bar{d}_{1} 
-\left(0.15\,\frac{\bar{g}_{0}}{\Fp}+0.28\,\frac{\bar{g}_{1}}{\Fp}
+0.01 \, \Fp^3 \bar{C}_{1}- 0.02\, \Fp^3 \bar{C}_{2}\right)
e\,\mathrm{fm} \ ,
\nonumber\\
\label{hEDMchiI}\\
d_{^{3}\mathrm{H}}(\mathrm{\chi I}) &=& 
0.85\,\bar{d}_{0}+0.95\,\bar{d}_{1} 
+\left(0.15\,\frac{\bar{g}_{0}}{\Fp}-0.28\,\frac{\bar{g}_{1}}{\Fp}
+0.01 \,\Fp^3 \bar{C}_{1}- 0.02\, \Fp^3 \bar{C}_{2}\right)e\,\mathrm{fm} \ .
\nonumber\\
\label{tEDMchiI}
\end{eqnarray}
{}From Eq. (\ref{NDATVCI}) we infer 
$\Fp\bar d_{0,1}/\bar g_0 =\Or(e\Fp/\mpi^2)\sim 2 \;e\,\mathrm{fm}$ and 
$\Fp^4 \bar{C}_{1,2}/\bar g_0 = \Or(\Fp^2/\mpi^2)\sim 2$. 
We see again that the $\bar g_{0,1}$ coefficients are somewhat smaller than 
expected; moreover, the $\slashPT$ short-range $N\!N$ interactions might
contribute even less.
However, one should keep in mind the large uncertainty in the $\bar{C}_{1,2}$ 
coefficients, and that these dimensional-analysis estimates could easily be
offset by dimensionless factors in the LECs.

\section{Discussion and Conclusions}
\label{discussion}

Historically, hadronic EDMs have mostly been discussed in the framework of
a one-boson-exchange model. It is assumed that $P$- and $T$-violation
is propagated by pions which are parametrized by three $\slashPT$
non-derivative interactions. In our notation,
\begin{eqnarray}
\mathcal{L} & = & -\frac{\bar{g}_{0}}{\Fp}\Nb\boldtau\cdot\boldpi N
-\frac{\bar{g}_{1}}{\Fp}\Nb\pi_{3}N
-\frac{\bar{g}_{2}}{\Fp}\Nb\tau_3\pi_{3}N
\end{eqnarray}
(in the nuclear physics literature, where chiral symmetry and power counting
are not emphasized, the coefficients are normally defined without $F_\pi$).
Hadronic EDMs are calculated as a function of these three parameters.
In some cases the effects of heavier bosons are included as well.
In this work we argue that this model is oversimplified. There is \textit{a
priori} no reason not to include $\slashPT$ photon-nucleon and 
short-range nucleon-nucleon interactions at low energies. 
By studying the chiral properties of
the fundamental $\slashPT$ sources of dimension up to six at the
QCD scale, it is possible to construct a model-independent hadronic
$\slashPT$ Lagrangian with a definite hierarchy between the different
$\slashPT$ hadronic interactions. It is found that the one-pion-exchange
model with three LECs is not appropriate for any of these
$\slashPT$ sources, and in general there are six
$\slashPT$ hadronic interactions that determine the EDMs of light
nuclei. Two of those are in the OPE model as well ---$\bar{g}_{0}$
and $\bar{g}_{1}$--- and the other four are additional interactions
that need to be considered when determining hadronic EDMs. 
The $\bar{g}_{2}$ interaction is not relevant at LO for any
of the fundamental sources. 
The other four necessary LECs 
are the isoscalar and isovector components of the neutron and proton EDMs
and two isoscalar $\slashPT$ $N\! N$ interactions of short range. 
The isovector $\slashPT$ $N\! N$ interactions come in at higher order for
all sources. 

We therefore propose that nuclear EDMs be analyzed
on the basis of these six LECs.
In the previous sections
we discussed  EDMs of light nuclei,
providing specific examples in the form of the deuteron,
helion, and triton.
In Table \ref{table1} the dependence of these various
EDMs on the six LECs is summarized. From the table it is clear that using
the OPE model gives an oversimplified 
view. At least six observables are required to identify the six LECs.
If other light nuclei become the target of experimental investigation,
their EDMs can be calculated along similar lines at the cost of 
larger computer resources. 
We hope that EDMs of heavier systems can be also expressed in terms of
these six LECs. However, in these cases there could be significant enhancement
factors for the $\slashPT$ potential contribution
\cite{HaxHenley83}, making important
otherwise subleading terms in the potential \cite{TVpotential},
such as the third non-derivative pion-nucleon coupling $\bar g_2$.

%%%%%%%%
%
\begin{table}
\caption{Dependence of the EDMs of the neutron, proton, deuteron,
helion, and triton on the six relevant $\slashPT$ low-energy constants. 
A ``-'' denotes that the LEC does not contribute 
in a model-independent way to the EDM at leading order. 
Values are for the Argonne $v18$ potential;
for the potential-model dependence of the results, see text.
} 
\begin{center}
\begin{tabular}{||c|cccccc||}
\hline 
LEC  
& $\bar{d}_{0}$  & $\bar{d}_{1}$  
& $(\bar{g}_{0}/\Fp)\, e\,\mathrm{fm}$  
& $(\bar{g}_{1}/\Fp)\, e\,\mathrm{fm}$
& $(\Fp^3\bar{C}_{1})\, e\,\mathrm{fm}$  
& $(\Fp^3\bar{C}_{2})\, e\,\mathrm{fm}$ \tabularnewline
\hline 
$d_{n}$  & $1$ & $-1$ & - & -  & - & -\tabularnewline
$d_{p}$  & $1$ & $1$  & - & -  & - & -\tabularnewline
$d_{^{2}\mathrm{H}}$  & $2$ & $0$  & $0.0002- 0.07\bt_1$  & $-0.19$  
& - & -\tabularnewline
$d_{^{3}\mathrm{He}}$ & $0.83$ & $-0.93$  & $-0.15$  & $-0.28$  
& $-0.01$ & $0.02$\tabularnewline
$d_{^{3}\mathrm{H}}$ & $0.85$ & $0.95$  & $0.15$  & $-0.28$  
& $0.01$ & $-0.02$\tabularnewline
\hline
\end{tabular}
\end{center}
\label{table1} 
\end{table}
%
%%%%%%%%

Once (a subset of) the LECs are determined it is possible to learn
something about the more fundamental $\slashPT$ sources at the QCD scale.
In Table \ref{table2}
we list for the different $\slashPT$ sources 
the expected orders of magnitude of the neutron EDM, $d_{n}$, 
and ratios between the other EDMs considered here and $d_{n}$.
Although some care is needed when using this table
---as we have discussed, the numbers found earlier
are not always exactly of the expected size---
it does allow some qualitative statements, even if 
less than six measurements are available.

\begin{table}
\caption{Expected orders of magnitude for the neutron EDM 
(in units of $e/\MQCD$),
the ratio of proton-to-neutron EDMs, the ratio of deuteron-to-neutron
EDMs, the ratio of helion-to-neutron EDMs, and
the ratio of triton-to-neutron EDMs, for the
$\tb$ term and the three dimension-six sources. $Q$ stands for the low-energy scales $F_\pi$, $m_\pi$, and $\gamma$.}
\begin{center}
\begin{tabular}{||c|cccc||}
\hline 
Source  & $\tb$  & qCEDM  & qEDM  & $\chi$I \tabularnewline
\hline 
$\MQCD\, d_{n}/e$  & $\Or\left(\tb\frac{\mpi^{2}}{\MQCD^{2}}\right)$  
& $\Or\left({\tilde{\delta}}\frac{\mpi^{2}}{M_{\slashT}^{2}}\right)$  
& $\Or\left(\delta\frac{\mpi^{2}}{M_{\slashT}^{2}}\right)$  
& $\Or\left(w\frac{\MQCD^{2}}{M_{\slashT}^{2}}\right)$ \tabularnewline
$d_{p}/d_{n}$  & $\Or\left(1\right)$  & $\Or\left(1\right)$  
& $\Or\left(1\right)$  & $\Or\left(1\right)$ \tabularnewline
$d_{^{2}\mathrm{H}}/d_{n}$  & $\Or\left(1\right)$  
& $\Or\left(\frac{\MQCD^{2}}{Q^2}\right)$  & $\Or(1)$  
& $\Or(1)$\tabularnewline
$d_{^{3}\mathrm{He}}/d_{n}$  & $\Or\left(\frac{\MQCD^{2}}{Q^2}\right)$  
& $\Or\left(\frac{\MQCD^{2}}{Q^2}\right)$  & $\Or(1)$  & $\Or(1)$
\tabularnewline
$d_{^{3}\mathrm{H}}/d_{n}$  & $\Or\left(\frac{\MQCD^{2}}{Q^2}\right)$  
& $\Or\left(\frac{\MQCD^{2}}{Q^2}\right)$  & $\Or(1)$  & $\Or(1)$
\tabularnewline
\hline
\end{tabular}
\end{center}
\label{table2} 
\end{table}

The simplest scenario is the one where $\slashPT$ is dominated by 
the qEDM, in which case all light nuclear EDMs are essentially given
by two LECs only: $\bar{d}_{0}$ and $\bar{d}_{1}$. 
(See Eqs. \eqref{dnLO}, \eqref{dpLO}, \eqref{dEDMqEDM},
\eqref{hEDMqEDM}, and \eqref{tEDMqEDM}.)
A measurement
of the proton and neutron EDMs would make deuteron and 
tri-nucleon EDMs 
testable predictions,
\begin{eqnarray}
d_{^2{\mathrm H}}&\simeq& d_n + d_p \ ,
\\
d_{^3{\mathrm He}}+d_{^3{\mathrm H}}&\simeq& 0.84 (d_n + d_p) \ ,
\\
d_{^3{\mathrm He}}-d_{^3{\mathrm H}}&\simeq& 0.94 (d_n - d_p) \ .
\end{eqnarray} 
The nucleon Schiff moments and the deuteron magnetic 
quadrupole moment (MQM) depend on other 
LECs \cite{Vri11a,Vri11b} and cannot be predicted.
For light nuclei the effects of the $\slashPT$ potential from the
qEDM are suppressed compared to the nucleon EDMs \cite{TVpotential},
although enhancements could make them more relevant for heavier nuclei.

$\slashPT$ from $\tb$ and qCEDM manifests itself in EDMs of light nuclei 
that differ significantly from the EDMs of their constituents. 
For both sources, the EDMs we calculated depend at LO
on four of the six LECs
---$\bar{g}_{0}$, $\bar{g}_{1}$, $\bar{d}_{0}$, and $\bar{d}_{1}$---
but in different ways.
For the qCEDM, the distinguishing feature
is that the deuteron EDM \eqref{dEDMqCEDM}
is expected to be significantly larger 
than the isoscalar nucleon EDM, thanks to $\bar g_1$.
Thus,
a measurement of nucleon and deuteron EDMs could be sufficient to 
qualitatively pinpoint, or exclude, qCEDM as a dominant $\slashPT$ source,
and to fix the values of $\bar d_{0,1}$ and $\bar g_1$. 
Then, the isoscalar combination of helion and triton EDMs, 
$d_{^3\mathrm{He}} + d_{^3\mathrm{H}}$, which in LO only depends on 
$\bar g_1$, becomes a falsifiable prediction of the theory,
\begin{equation}
d_{{}^3{\mathrm {He}}}+d_{{}^3{\mathrm {H}}}\simeq 3 d_{{}^2{\mathrm H}} \ .
\label{careful}
\end{equation}
If, to be on the safe side, we keep some subleading terms (the nucleon EDMs) as we did in Eqs. (\ref{hEDMqCEDM}) and (\ref{tEDMqCEDM}), then we get (including the subleading term  $2 \bar{d}_0$ in Eq. (\ref{dEDMqCEDM})) an additional $- 2.16 (d_n + d_p)$ on the right-hand side of Eq. (\ref{careful}). Furthermore, $\bar g_0$ can then be extracted from 
$d_{^3\mathrm{He}} - d_{^3\mathrm{H}}$ 
(see Eqs. (\ref{hEDMqCEDM}) and (\ref{tEDMqCEDM})), 
leading to testable predictions for other $\slashPT$ observables.

In contrast, for the Standard
Model $\tb$ term we do not expect the deuteron EDM to be significantly 
different from twice the isoscalar nucleon EDM. 
Although the deuteron EDM \eqref{dEDMtheta}
formally depends on the isoscalar nucleon EDM and on the pion-nucleon couplings $\bar g_0$ and $\bar g_1$, the results of 
Section \ref{Deuteron} show that the pion-exchange contribution is 
likely only $\sim 10$\% of the nucleon EDM. 
On the other hand, the EDMs of $^3$He and $^3$H, Eqs. 
\eqref{hEDMtheta} and \eqref{tEDMtheta},
are dominated by $\bar g_0$, although
they receive important contributions from the neutron and proton EDMs. 
In particular, we expect the isovector combination 
$d_{^3\mathrm{He}} - d_{^3\mathrm{H}}$, which is sensitive to $\bar g_0$, 
to differ from the isovector nucleon EDM $\bar d_1$, 
while the isoscalar combination $d_{^3\mathrm{He}}+ d_{^3\mathrm{H}}$ 
should be close to $2 \bar d_0$:
\begin{eqnarray}
d_{^3{\mathrm He}}+d_{^3{\mathrm H}}&\simeq& 0.84 (d_n + d_p) \ ,
\\
d_{^3{\mathrm He}}-d_{^3{\mathrm H}}&\ne& 0.94 (d_n - d_p) \ .
\end{eqnarray}
The experimental observation of these relations in nucleon, deuteron, helion, 
and triton EDM experiments would qualitatively indicate the $\tb$
term as the main source responsible for $\slashPT$. 
Quantitatively, the measurement of nucleon, helion, and triton EDM allows 
extraction of the coupling $\bar g_0$, 
which then can be used to provide testable predictions of other $\slashPT$  
observables, like the proton Schiff moment \cite{Vri11a,Thomas} or 
the deuteron MQM \cite{Vri11b}, which are not sensitive to the nucleon EDMs.

Finally, in the case of the $\chi$I sources the analysis is in principle 
most complicated, due to the appearance of all six LECs.
Like for $\tb$, the deuteron EDM \eqref{dEDMchiI},
although formally dependent on $\bar{g}_{1}$ at LO, is probably
dominated by $\bar{d}_{0}$. 
The tri-nucleon EDMs \eqref{hEDMchiI} and \eqref{tEDMchiI} 
formally depend on all six LECs, but they are again possibly
dominated by $\bar d_0$ and $\bar d_1$. 
It might thus be difficult to separate the $\chi$I sources from qEDM. 
For less dilute, but still light, systems we expect different results. 
For these systems, in the case of qEDM the EDMs are still dominated 
by $\bar d_{0,1}$, but for $\chi$I sources 
we expect the contributions from the $\slashPT$ potential to 
be more significant, implying that measurements on these systems might 
separate $\chi$I sources from qEDM. 
Of course, more extensive calculations are necessary to verify this claim.

In conclusion, we have argued that an experimental program
to measure light nuclear EDMs could offer valuable information
on yet undiscovered sources of parity and time-reversal violation.
Our case is based on some crucial, but relatively general
assumptions, such as the validity of the Standard Model with its
minimal particle content at the electroweak scale,
and the naturalness of interaction strengths.
Elsewhere \cite{Vri11b}
it has already been pointed out ---basically on the basis of
dimensional analysis--- that sensitivity to the deuteron
EDM at the level hoped for in storage ring experiments \cite{storageringexpts}
would probe scales where new physics is expected.
A similar analysis holds for our tri-nucleon results, 
Eqs. \eqref{helionEDM} and \eqref{tritonEDM}.
But our results here go beyond dimensional analysis
and suggest 
that, at least for the lightest 
nuclei, the contribution of the neutron and proton EDMs are more important 
than expected by simple power counting. 
For all sources, they compete with, 
when they do not dominate, the effects of the $\slashPT$ potential. 
For this reason,  other $\slashPT$ observables insensitive to the nucleon EDMs,
for example higher $\slashPT$ electromagnetic moments, could provide important 
complementary information and a cleaner way to extract pion-nucleon and 
nucleon-nucleon $\slashPT$ couplings.
Additionally, it would be interesting if EDMs of heavier
systems could be recast in terms of our EFT approach.

\section*{Acknowledgments}

We thank the organizers of the Workshop on Search for Electric Dipole
Moments at Storage Rings (Physikzentrum Bad Honnef, July 2011),
H. Str\"oher and F. Rathmann, for providing a stimulating atmosphere for
discussions with many colleagues, in particular  J. Bsaisou, 
B. Gibson, C. Hanhart, A. Nogga, G. Onderwater, and A. Wirzba.  
We are particularly grateful to P. Navr\'atil for the NCSM code
of the $PT$ wavefunctions, and to C. Hanhart for
comments on the manuscript.
C.-P. Liu and U. van Kolck acknowledge the hospitality of KVI, 
where this research was carried out. 
This research was supported  by the Dutch Stichting FOM
under programs 104 and 114 (JdV, RH, RGET),  by the ROC NSC under
grant NSC98-2112-M-259-004-MY3 (CPL), and by the US DOE under
grants DE-FG02-06ER41449 (EM), DE-FG02-04ER41338 (EM, UvK),
and DE-FC02-07ER41457 (IS). The three-body calculations were
performed on the University of Washington Hyak cluster
(NSF MRI grant PHY-0922770).

\appendix

\section{Potential in Coordinate Space}
\label{Vcoord}

In configuration space, the LO potential of Sect. \ref{LOpotmom}
is given by \cite{TVpotential} 
\begin{eqnarray}
V_{\slashPTsub}(\vec{r}\,) & = & -\frac{\bar{g}_{0}g_{A}}{F_{\pi}^{2}}
\boldtau^{(i)}\cdot\boldtau^{(j)}
\left(\vec{\sigma}^{\,(i)}-\vec{\sigma}^{\,(j)}\right)
\cdot\left(\vec{\nabla}_{r}\, U(r)\right)
\nonumber \\
& & -\frac{\bar{g}_{1}g_{A}}{2F_{\pi}^{2}}
\left[\left(\tau_{3}^{(i)}+\tau_{3}^{(j)}\right)
\left(\vec{\sigma}^{(i)}-\vec{\sigma}^{(j)}\right)
+\left(\tau_{3}^{(i)}-\tau_{3}^{(j)}\right)
\left(\vec{\sigma}^{(i)}+\vec{\sigma}^{(j)}\right)\right]
\cdot\left(\vec{\nabla}_{r}U(r)\right)
\nonumber \\
& & +\frac{1}{2}
\left[\bar{C}_{1}+\bar{C}_{2}\boldtau^{(i)}\cdot\boldtau^{(j)}\right]
\left(\vec{\sigma}^{\,(i)}-\vec{\sigma}^{\,(j)}\right)\cdot
\left(\vec{\nabla}_{r}\,\delta^{(3)}(\vec{r}\,)\right)\ ,
\end{eqnarray}
where $\vec{r}=\vec x_i-\vec x_j$ is the relative position of the 
two interacting
nucleons and \begin{equation}
U(r)=\frac{1}{12\pi r}\left[2\exp\left(-m_{\pi^{\pm}}r\right)
+\exp\left(-m_{\pi^{0}}r\right)\right]\ ,\label{Yukawa}
\end{equation}
which reduces to the usual Yukawa function $U(r)=\exp(-m_{\pi}r)/4\pi r$
when, at LO, we ignore the pion mass difference.

Analogously, the NNLO potential of Sect. \ref{subLOpot} 
becomes \cite{TVpotential}
\begin{eqnarray}
V_{\slashPTsub}(\vec{r}, \vec{\nabla}_r , \vec{\nabla}_X\,)  & = &    
-\frac{\bar g_0 g_A}{2 F^2_{\pi}}
\left[\left(\frac{\bar g_1}{\bar g_0} - \frac{\beta_1}{2 g_A}\right)  
\left(\tau_3^{(i)} + \tau_3^{(j)} \right) \,
\left( \vec\sigma^{(i)}  -  \vec\sigma^{(j)}\right) 
\right.
\nonumber \\
&&\left. + 
\left(\frac{ \bar g_1}{\bar g_0} + \frac{\beta_1}{2 g_A}\right) 
\left(\tau_3^{(i)} - \tau_3^{(j)} \right) \,
\left( \vec\sigma^{(i)} +  \vec\sigma^{(j)}\right) 
 \right] \cdot \left(\vec  \nabla_r U(r) \right)
\nonumber\\
 &&   
+\frac{ \bar g_0 g_A}{3 F^2_{\pi}}    
\left(3\tau^{(i)}_3\tau^{(j)}_3 -\boldtau^{(i)}\cdot\boldtau^{(j)}\right)
\nonumber\\
&&\times
\left(\vec\sigma^{(i)} - \vec\sigma^{(j)}\right) \cdot 
\left[\frac{ \delta m_N^2  }{2 m_{\pi}} \left(\vec \nabla_r \; r U(r) \right) 
+ \left(\vec \nabla_r  W(r)  \right)\right] 
%,
\nonumber\\
&&
-i\frac{\bar g_0 g_A}{2 F^2_{\pi}} \frac{\delta m_N}{m_N}
\left(\tau^{(i)}\times \tau^{(j)}\right)_3
\left\{\left(\vec\sigma^{(i)} + \vec\sigma^{(j)}\right) \cdot 
\left\{\vec{\nabla}_r ,U(r)\right\}
\right.
\nonumber\\
&&\left.
+\left(\vec\sigma^{(i)} - \vec\sigma^{(j)}\right) \cdot 
\left[U(r)\vec{\nabla}_X-\frac{1}{m_\pi}
\left(\vec{\nabla}_r \nabla_r^n rU(r)\right) \nabla_X^n \right] 
\right\}
,   
\end{eqnarray}
where $\vec X=(\vec x_i +\vec x_j)/2$ and
\begin{equation}
W(r) =  \frac{1}{4\pi r} 
\left[\exp{(-m_{\pi^{\pm} } r)}- \exp{(-m_{\pi^0} r)}\right],
\label{Yukawadiff}
\end{equation}
which is entirely a consequence of isospin breaking.

\section{Fourier Transform of the Currents}
\label{Fourier}

To evaluate the matrix elements in Section \ref{Deuteron} we need
to transform the currents to configuration space. 
We follow Ref. \cite{Koelling:2011mt}
and transform with respect to the nucleon momenta but not with respect
to the photon momentum. In the most general case 
\begin{eqnarray}
J^{0}(\vec{x}_{i},\vec{x}_{i}^{\; '},\vec{x}_{j},\vec{x}_{j}^{\; '},\vec{q}\,) 
& = & 
\int\frac{\mathrm{d}^{3}p_{i}}{(2\pi)^{3}}
\int\frac{\mathrm{d}^{3}p'_{i}}{(2\pi)^{3}}
\int\frac{\mathrm{d}^{3}p_{j}}{(2\pi)^{3}}
\int\frac{\mathrm{d}^{3}p'_{j}}{(2\pi)^{3}}
e^{-i\vec{p}_{i}\cdot\vec{x}_{i}}e^{-i\vec{p}_{j}\cdot\vec{x}_{j}}
e^{i\vec{p}_{i}^{\; '}\cdot\vec{x}_{i}^{\; '}}
e^{i\vec{p}_{j}^{\; '}\cdot\vec{x}_{j}^{\; '}}
\nonumber \\
& & (2\pi)^{3}
\delta^{(3)}(\vec{p}_{i}+\vec{p}_{j}-\vec{p}_{i}^{\; '}
-\vec{p}_{j}^{\; '}-\vec{q}\,)
J^{0}(\vec{p}_{i},\vec{p}_{i}^{\; '},\vec{p}_{j},\vec{p}_{j}^{\; '},\vec{q}\,)
\ .
\end{eqnarray}
Introducing the relative configuration-space coordinates 
$\vec{r}=\vec{x}_{i}-\vec{x}_{j}$,
$\vec{r}\,'=\vec{x}_{i}'-\vec{x}_{j}'$, 
$\vec{X}=(\vec{x}_{i}+\vec{x}_{j})/2$,
and $\vec{X}'=(\vec{x}_{i}^{\; '}+\vec{x}_{j}^{\; '})/2$, 
we rewrite this as 
\begin{eqnarray}
J^{0}(\vec{r},\vec{r}\,',\vec{X},\vec{X}',\vec{q}\,) & = & 
e^{-\frac{i}{2}\vec{q}\cdot(\vec{X}+\vec{X}')}
\int\frac{\mathrm{d}^{3}P_{t}}{(2\pi)^{3}}
\int\frac{\mathrm{d}^{3}K}{(2\pi)^{3}}
\int\frac{\mathrm{d}^{3}k}{(2\pi)^{3}}
\nonumber \\
& & e^{-i\vec{P}_{t}\cdot(\vec{X}-\vec{X}')}
e^{-i\vec{K}\cdot(\vec{r}-\vec{r}\,')}
e^{-\frac{i}{2}\vec{k}\cdot(\vec{r}+\vec{r}\,')}
J^{0}(\vec{q},\vec{k},\vec{K},\vec{P}_{t})\ .
\end{eqnarray}

The currents we need (the third currents in Eqs. (\ref{eq:J0_PCTC_2B})
and (\ref{eq:J0_PVTV_2B})) depend on $\vec{q}$ and $\vec{k}$ only,
such that the expression can be simplified to 
\begin{eqnarray}
J^{0}(\vec{r},\vec{X},\vec{q}\,) & = &  e^{-i\vec{q}\cdot\vec{X}}
\int\frac{\mathrm{d}^{3}k}{(2\pi)^{3}}e^{-i\vec{k}\cdot\vec{r}}
J^{0}(\vec{q},\vec{k}\,)\ .\end{eqnarray}
The Fourier transforms can be done and we find for the required currents
\begin{eqnarray}
J_{PT,c}^{0}(\vec{r},\vec{X},\vec{q}\,) & = & 
-\frac{2eg_{A}^{2}}{\Fp^{2}}\delta m_{N}
\left(\boldtau^{(i)}\cdot\boldtau^{(j)}-\tau_{3}^{(i)}\tau_{3}^{(j)}\right) 
\nonumber \\ 
&& \times
e^{-i\vec{q}\cdot\vec{X}}
\left[\vec{\sigma}^{(i)}\cdot\left(i\vec{\nabla}_{r}+\frac{\vec{q}}{2}\right)
\vec{\sigma}^{(j)}\cdot\left(i\vec{\nabla}_{r}-\frac{\vec{q}}{2}\right)
\right]W(\vec{q},\vec{r}\,)\ ,\\
J_{\slashPTsub,c}^{0}(\vec{r},\vec{X},\vec{q}\,) & = & 
 \frac{2ieg_{A}\bar{g}_{0}}{\Fp^{2}}\delta m_{N}
\left(\boldtau^{(i)}\cdot\boldtau^{(j)}-\tau_{3}^{(i)}\tau_{3}^{(j)}\right) 
\nonumber \\ 
&& \times
e^{-i\vec{q}\cdot\vec{X}}\left[
\left(\vec{\sigma}^{(i)}+\vec{\sigma}^{(j)}\right)\cdot\frac{\vec{q}}{2}
+\left(\vec{\sigma}^{(i)}-\vec{\sigma}^{(j)}\right)\cdot
\left(i\vec{\nabla}_{r}\right)\right]W(\vec{q},\vec{r}\,)\ ,
\end{eqnarray}
in terms of the function 
\begin{eqnarray}
W(\vec{q},\vec{r}\,) & = & \frac{1}{8\pi}\int_{0}^{1}\mathrm{d}\alpha\;
\mathrm{exp}\left[i\frac{\vec{q}\cdot\vec{r}}{2}(1-2\alpha)\right]
\frac{\mathrm{exp}[-r(\mpi^{2}+\vec{q}^{\,2}\alpha(1-\alpha))^{1/2}]}
{[\mpi^{2}+\vec{q}^{\,2}\alpha(1-\alpha)]^{1/2}}\ .
\end{eqnarray}

Before continuing it is convenient to look at the inverse Fourier
transform of the current 
\begin{eqnarray}
J^{0}(\vec{q}\,) & = & 
\int\mathrm{d}^{3}x\; e^{-i\vec{q}\cdot\vec{x}}J^{0}(\vec{x}\,)
\nonumber \\
& = & \int\mathrm{d}^{3}x\; J^{0}(\vec{x})
-i\vec{q}\cdot\int\mathrm{d}^{3}x\;\vec{x}J^{0}(\vec{x})
+\Or(\vec{q}^{\,2})
\nonumber \\
& = & Z e-i\vec{q}\cdot\vec{D}+\Or(\vec{q}^{\,2})\ ,
\end{eqnarray}
where $Ze$ is the total charge and $\vec{D}$ is the EDM operator
used in Sections \ref{Deuteron} and \ref{helium}. 
An easy way to extract the EDM operator
is by using 
\begin{eqnarray}
\vec{D} & = & i\lim_{q\rightarrow0}\vec{\nabla}_{q}J^{0}(\vec{q}\,)\ .
\label{currenttoEDM}
\end{eqnarray}

As an example we consider the EDM operator coming from 
$J_{\slashPTsub,c}^{0}(\vec{r},\vec{X},\vec{q}\,)$. 
{}From Eq. \eqref{currenttoEDM} we read off
\begin{eqnarray}
\vec{D}_{\slashPTsub,c} & = & \frac{2ieg_{A}\bar{g}_{0}}{\Fp^{2}}\delta m_{N}
\left(\boldtau^{(i)}\cdot\boldtau^{(j)}-\tau_{3}^{(i)}\tau_{3}^{(j)}\right) 
\nonumber \\ 
&& \times
\left[\frac{i}{2}\left(\vec{\sigma}^{(i)}+\vec{\sigma}^{(j)}\right)
+ \vec X \left(\vec{\sigma}^{(i)}-\vec{\sigma}^{(j)}\right)
\cdot\left(i\vec{\nabla}_{r}\right)\right]
\frac{e^{- \mpi\, r}}{8\pi\mpi}
\nonumber\\
&= & -\frac{eg_{A}\bar{g}_{0}}{\Fp^{2}}\delta m_{N}
\left(\boldtau^{(i)}\cdot\boldtau^{(j)}-\tau_{3}^{(i)}\tau_{3}^{(j)}\right) 
\nonumber \\ 
&& \times
\left[\left(\vec \sigma^{(i)}\cdot\vec \nabla^{(i)}
+\vec \sigma^{(j)}\cdot\vec \nabla^{(j)}\right)\left(\vec x_i + \vec x_j\right)
\frac{e^{- \mpi\, \abs{\vec x_i-\vec x_j}}}{8\pi\mpi} \right],
\end{eqnarray}
where we used $ \lim_{q\rightarrow0}\vec{\nabla}_{q} W(\vec q, \vec r)=0$.
This is Eq. \eqref{D2slashPT}. 
Following similar steps we obtain Eq. \eqref{D2PT} from 
$J_{PT,c}^{0}(\vec{r},\vec{X},\vec{q}\,)$.


\begin{thebibliography}{64}

\bibitem{KhripLam1997}
I. B. Khriplovich and S. K. Lamoreaux,
\textit{CP Violation Without Strangeness: Electric Dipole Moments
of Particles, Atoms, and Molecules\/} (Springer Verlag, Berlin, 1997).

\bibitem{Pospelov:2005pr}
M. Pospelov and A. Ritz,
Ann. Phys. {\bf 318}, 119 (2005).

\bibitem{Kobayashi:1973fv}
M. Kobayashi and T. Maskawa,
Prog. Theor. Phys. {\bf 49}, 652 (1973).

\bibitem{expts}
T. M. Ito, 
J. Phys. Conf. Ser. {\bf 69}, 012037 (2007), {\tt nucl-ex/0702024};
K. Bodek {\it et al.}, 
{\tt arXiv:0806.4837 [nucl-ex]}.

\bibitem{dnbound}
C. A. Baker {\it et al.}, 
Phys. Rev. Lett. {\bf 97}, 131801 (2006).

\bibitem{storageringexpts}
F. J. M. Farley {\it et al.},
Phys. Rev. Lett. {\bf 93}, 052001 (2004);
Y. F. Orlov, W. M. Morse, and Y. K. Semertzidis,
Phys. Rev. Lett. {\bf 96}, 214802 (2006);
C. J. G. Onderwater, 
J. Phys. Conf. Ser. {\bf 295}, 012008 (2011).

\bibitem{hgbound}
W. C. Griffith {\it et al.}, 
Phys. Rev. Lett. {\bf 102}, 101601 (2009).

\bibitem{McKellar:1987tfPospelov:1994uf}
I. B. Khriplovich and A. R. Zhitnitsky,
Phys.\ Lett.\  B {\bf 109}, 490 (1982);
B. H. J. McKellar, S. R. Choudhury, X.-G. He, and S. Pakvasa,
Phys.\ Lett.\  B {\bf 197}, 556 (1987);
X.-G. He, B. H. J. McKellar, and S. Pakvasa,
Int. J. Mod. Phys. {\bf A4}, 5011 (1989);
{\bf A6}, 1063(E) (1991);
M. E. Pospelov,
Phys.\ Lett.\  B {\bf 328}, 441 (1994);
A. Czarnecki and B. Krause,
Phys.\ Rev.\ Lett.\ {\bf 78}, 4339 (1997).

\bibitem{'tHooft:1976up}
G. 't Hooft,
Phys. Rev. Lett. {\bf 37}, 8 (1976);
C. G. Callan, Jr, R. F. Dashen, and D. J. Gross,
Phys. Lett. B \textbf{63}, 334 (1976);
R. Jackiw and C. Rebbi,
Phys. Rev. Lett. {\bf 37}, 172 (1976).

\bibitem{Bal79}
V. Baluni, 
Phys. Rev. D {\bf 19}, 2227 (1979).

\bibitem{Cre79}
R. J. Crewther, P. Di Vecchia, G. Veneziano, and E. Witten,
Phys. Lett. B {\bf 88}, 123 (1979); {\bf 91}, 487(E) (1980).

\bibitem{dim6origin}
W.~Buchm\"uller and D.~Wyler,
Nucl. Phys. {\bf B268}, 621 (1986);
A.~De R\'ujula, M. B.~Gavela, O.~P\`ene, and F. J.~Vegas,
Nucl. Phys. {\bf B357}, 311 (1991).

\bibitem{Weinberg:1989dx}
S. Weinberg,
Phys. Rev. Lett. {\bf 63}, 2333 (1989).

\bibitem{Grzadkowski:2010es}
B.~Grzadkowski, M.~Iskrzynski, M.~Misiak, and J.~Rosiek,
JHEP {\bf 1010}, 085 (2010).

\bibitem{RamseyMusolf:2006vr}
M. J.~Ramsey-Musolf and S.~Su,
Phys. Rept. {\bf 456}, 1 (2008).

\bibitem{weinberg79} 
S. Weinberg, Physica {\bf 96A}, 327 (1979);
J. Gasser and H. Leutwyler, Ann. Phys. {\bf 158}, 142 (1984);
Nucl. Phys. {\bf B250}, 465 (1985).

\bibitem{HBChPT} 
E. Jenkins and A. V. Manohar,
Phys. Lett. B {\bf 255}, 558 (1991).

\bibitem{Weinbergbook} 
S. Weinberg, 
\textit{The Quantum Theory of Fields\/}, Vol. 2
(Cambridge University Press, Cambridge, 1996).

\bibitem{ulfreview}
V. Bernard, N. Kaiser, and U.-G Mei{\ss}ner,
Int. J. Mod. Phys. E {\bf 4}, 193 (1995).

\bibitem{Thomas}
S. D. Thomas,
Phys. Rev.  D {\bf 51}, 3955 (1995).

\bibitem{BiraHockings} 
W. H. Hockings and U. van Kolck, 
Phys. Lett. B {\bf 605}, 273 (2005).

\bibitem{BiraEmanuele} 
E. Mereghetti, W. H. Hockings, and U. van Kolck,
Ann. Phys. {\bf 325}, 2363 (2010).

\bibitem{Mer11}
E. Mereghetti, J. de Vries, W. H. Hockings, C. M. Maekawa, and U. van Kolck,
Phys. Lett. B {\bf 696}, 97 (2011).

\bibitem{su3}
H.-Y. Cheng, 
Phys. Rev. D {\bf 44}, 166 (1991);
A. Pich and E. de Rafael, 
Nucl. Phys. {\bf B367}, 313 (1991);
P. Cho, 
Phys. Rev. D {\bf 48}, 3304 (1993);
B. Borasoy, 
Phys. Rev. D {\bf 61}, 114017 (2000).

\bibitem{Ottnad}
S. Narison, 
Phys. Lett. B {\bf 666}, 455 (2008);
K. Ottnad, B. Kubis, U.-G. Mei{\ss}ner, and F.-K. Guo,
Phys. Lett. B {\bf 687}, 42 (2010).

\bibitem{Vri11a}
J. de Vries, E. Mereghetti, R. G. E. Timmermans, and U. van Kolck,
Phys. Lett. B {\bf 695}, 268 (2011).

\bibitem{dim6} 
J. de Vries, E. Mereghetti, R. G. E. Timmermans, and U. van Kolck,
in preparation.

\bibitem{Vri11b}
J. de Vries, E. Mereghetti, R.G.E. Timmermans, and U. van Kolck,
Phys. Rev. Lett. {\bf 107}, 091804 (2011).

\bibitem{KSW}
D. B. Kaplan, M. J. Savage, and M. B. Wise,
Nucl. Phys. {\bf B534}, 329 (1998);
S. Fleming, T. Mehen, and I.W. Stewart, 
Nucl. Phys. {\bf A677}, 313 (2000).

\bibitem{nEFT}
P. F. Bedaque and U. van Kolck,
Ann. Rev. Nucl. Part. Sci. \textbf{52}, 339 (2002);

\bibitem{weinberg}
S. Weinberg, 
Nucl. Phys. B \textbf{363}, 3 (1991).

\bibitem{vanKolck}
U. van Kolck, 
Ph.D. dissertation, University of Texas (1993);
Few-Body Syst. Suppl.  {\bf 9}, 444 (1995).

\bibitem{morethanweinberg}
S. R. Beane, P. F. Bedaque, M. J. Savage, and U. van Kolck,
Nucl. Phys. A \textbf{700}, 377 (2002);
A. Nogga, R. G. E. Timmermans, and U. van Kolck,
Phys. Rev. C \textbf{72}, 054006 (2005);
M. C. Birse,
Phys. Rev. C {\bf 74}, 014003 (2006);
{\bf 76}, 034002 (2007);
M. Pav\'on Valderrama,
Phys. Rev. C {\bf 83}, 024003 (2011);
{\tt arXiv:1108.0872 [nucl-th]};
Bingwei Long and C.-J. Yang,
{\tt arXiv:1108.0985 [nucl-th]}.

\bibitem{TVpotential}
C. M. Maekawa, E. Mereghetti, J. de Vries, and U. van Kolck,
{\tt arXiv:1106.6119 [nucl-th]}.

\bibitem{juelich}
J. Bsaisou, C. Hanhart, S. Liebig, A. Nogga, and A. Wirzba,
talk at the Workshop on Search for Electric Dipole Moments
at Storage Rings, Physikzentrum Bad Honnef, July 2011,
www2.fz-juelich.de/ikp/edm/en/abstract.php?IDABS=28. 

\bibitem{SFK84}
O. P. Sushkov, V. V. Flambaum, and I. B. Khriplovich,
Sov. Phys. JETP \textbf{60}, 873 (1984).

\bibitem{Avi85}
Y. Avishai,
Phys. Rev. D \textbf{32}, 314 (1985).

\bibitem{Khrip}
I. B. Khriplovich and R. V. Korkin, 
Nucl. Phys. {\bf A665}, 365 (2000).

\bibitem{Liu04}
C.-P. Liu and R. G. E. Timmermans, 
Phys. Rev. C {\bf 70}, 055501 (2004).

\bibitem{Afn10} 
I. R. Afnan and B. F. Gibson, 
Phys. Rev. C \textbf{82}, 064002 (2010).

\bibitem{Avi86}
Y. Avishai and M. Fabre de la Ripelle, 
Phys. Rev. Lett. {\bf 56}, 2121 (1986);
Nucl. Phys. {\bf A468}, 578 (1987).

\bibitem{Ste08}
I. Stetcu, C.-P. Liu, J. L. Friar, A. C. Hayes, and P. Navr\'{a}til, 
Phys. Lett. B {\bf 665}, 168 (2008).

\bibitem{Her66}
P. Herczeg, 
Nucl. Phys. {\bf 75}, 655 (1966);
C.-P. Liu and R. G. E. Timmermans,
Phys. Lett. B {\bf 634}, 488 (2006).

\bibitem{Sto94}
V. G. J. Stoks, R. A. M. Klomp, C. P. F. Terheggen, and J. J. de Swart,
Phys. Rev. C {\bf 49}, 2950 (1994).

\bibitem{Wir95}
R. B. Wiringa, V. G. J. Stoks, and R. Schiavilla,
Phys. Rev. C {\bf 51}, 38 (1995).

\bibitem{eftinspiredNN}
D. R. Entem and R. Machleidt,
Phys. Rev. C {\bf 68}, 041001(R) (2003);
E. Epelbaum, A. Nogga, W. Gl\"ockle, H. Kamada, U.G. Mei{\ss}ner, and
H. Wita\l a,
Phys. Rev. C {\bf 66}, 064001 (2002).

\bibitem{hybrid}
S. Weinberg, 
Phys. Lett. B \textbf{295}, 114 (1992).

\bibitem{dan}
D. R. Phillips, 
Phys. Lett. B {\bf 567}, 12 (2003).

\bibitem{CPonP}
C.-P. Liu,
Phys. Rev. C {\bf 75}, 065501 (2007).

\bibitem{HaxHenley83}  
W.C. Haxton and E.M. Henley,
Phys. Rev. Lett. \textbf{51}, 1937 (1983).

\bibitem{NDA}
A. Manohar and H. Georgi,
Nucl. Phys. \textbf{B234}, 189 (1984).

\bibitem{Bernard:1992qa}
V.~Bernard, N.~Kaiser, J.~Kambor, and U.-G.~Mei{\ss}ner,
Nucl. Phys. {\bf B388}, 315 (1992);
N.~Fettes, U.-G.~Mei{\ss}ner, and S.~Steininger,
Nucl. Phys. {\bf A640}, 199 (1998).

\bibitem{Friar:2004ca}
J. L. Friar, U. van Kolck, M. C. M. Rentmeester, and R. G. E. Timmermans,
Phys. Rev.  C {\bf 70}, 044001 (2004).

\bibitem{Nakamura:2010zzi}
K.~Nakamura  [Particle Data Group],
J. Phys. G {\bf 37}, 075021 (2010).

\bibitem{latticedeltamN}
S. R. Beane, K. Orginos, and M. J. Savage,
Nucl. Phys. {\bf B768}, 38 (2007).

\bibitem{CSBd}
U. van Kolck, J. A. Niskanen, and G. A. Miller,
Phys. Lett. B {\bf 493}, 65 (2000);
D. R. Bolton and G. A. Miller,
Phys. Rev. C {\bf 81}, 014001 (2010);
A. Filin {\it et al.},
Phys. Lett. B {\bf 681}, 423 (2009).

\bibitem{Cott}
J. Gasser and H. Leutwyler,
Phys. Rep. {\bf 87}, 77 (1982).

\bibitem{piNcoupling}
V. G. J. Stoks, R. Timmermans, and J. J. de Swart,
Phys. Rev.  C {\bf 47},  512 (1993);
M. C. M. Rentmeester, R. G. E. Timmermans, J.L. Friar, and J. J. de Swart,
Phys. Rev. Lett.  {\bf 82},  4992 (1999).

\bibitem{isoviolphen}
U. van Kolck, J. L. Friar, and T. Goldman,
Phys. Lett. B \textbf{371}, 169 (1996);
U. van Kolck, M. C. M. Rentmeester, J. L. Friar, T. Goldman, 
and J. J. de Swart,
Phys. Rev. Lett. {\bf 80}, 4386 (1998).

\bibitem{Manohar}
A. V. Manohar,
Phys. Rev.  D {\bf 56}, 230 (1997).

\bibitem{Liebig:2010ki}
S. Liebig, V. Baru, F. Ballout, C. Hanhart, A. Nogga,
Eur. Phys. J.  {\bf A47}, 69 (2011).


\bibitem{NCSM}
P. Navr\'atil, J. P. Vary, and B. R. Barrett,
Phys. Rev. Lett. {\bf 84}, 5728 (2000);
Phys. Rev.  C {\bf 62},  054311 (2000).

\bibitem{lithuania}
P. Navr\'atil, G. Kamuntavicius, and B. R. Barrett,
Phys. Rev.  C {\bf 61},  044001 (2000).

\bibitem{Koelling:2011mt}
S. K\"olling, E. Epelbaum, H.~Krebs, and U.-G. Mei{\ss}ner,
{\tt arXiv:1107.0602 [nucl-th]}.  

\end{thebibliography}
\end{document}